\documentclass{aa}
\usepackage{graphicx}
\usepackage{txfonts}
\usepackage{xcolor}

\begin{document}

   \title{An empirical metallicity tracer in CEMP and C-normal stars.\thanks{Based on observations obtained at ESO Paranal Observatory, programmes 084.D-0117(A), 085.D-0041(A), and  090.D-0321(A), see also acknowledgements.}}

   \author{D. Singh\inst{1,2}
          \and
          C. J. Hansen\inst{2,1}
          \and
          J. S. Byrgesen\inst{1,2}
          \and
          M. Reichert\inst{3}
          \and
          H. M. Reggiani\inst{4}
          }

   \institute{University of Copenhagen, Dark Cosmology Centre, The Niels Bohr Institute, 
Vibenshuset, Lyngbyvej 2, DK-2100 Copenhagen, Denmark\\
              \email{danielzing@hotmail.com}
         \and
             Max Planck Institute for Astronomy, K\"onigstuhl 17, D-69117 Heidelberg, Germany\\
             \email{hansen@mpia.de}
        \and
        Technische Universit\"at Darmstadt, Institut f\"ur Kernphysik, Schlossgartenstr. 2, 64289 Darmstadt, Germany
        \and
        Department of Physics and Astronomy, Johns Hopkins University, 3400 N Charles St., Baltimore, MD 21218\\
        \email{hreggiani@gmail.com}
        }

   \date{}

 \abstract
   {Deriving the metallicity, [Fe/H], in low-resolution spectra of carbon-enhanced metal-poor (CEMP) stars is a tedious task that, owing to the large number of line blends, often leads to uncertainties on [Fe/H] exceeding 0.25\,dex. The CEMP stars increase in number with decreasing [Fe/H] and some of these are known to be bona fide second generation halo stars. Hence, knowing their [Fe/H] is important for tracing the formation and chemical evolution of the Galaxy. }
   {Here, we aim to improve the [Fe/H] measurements in low-resolution spectra by avoiding issues related to blends. In turn, we improve our chemical tagging in such spectra at low metallicities.}
   {We developed an empirical way of deriving [Fe/H] in CEMP (and C-normal) stars that relates the equivalent width (EW) of strong lines, which remain detectable in lower-resolution, metal-poor spectra, such as X-Shooter spectra to [Fe/H]. }
   {The best [Fe/H] tracers are found to be Cr I and Ni I, which both show strong transitions in spectral regions that are free of molecular bands (between $\sim 5200-6800$\AA\,, a region accessible to most surveys). We derive different relations for dwarfs and giants. The relations are valid in the ranges $\sim$  $-3<$[Fe/H]$<-0.5$ and $10<$EW$<800$\,m\AA\ (Cr) or [Fe/H]$>-3.2$ and EW$>5$\,m\AA\ (Ni), depending on the trace element and line as well as the stellar evolutionary stage. }
   {The empirical relations are valid for both CEMP and C-normal stars and have been proven to be accurate tracers in a sample of $\sim 400$ stars (mainly giants). The metallicities are accurate to within $\pm \sim 0.2$ dex depending on the sample and resolution, and the empirical relations are robust to within 0.05-0.1\,dex. Our relations will improve the metallicity determination in future surveys, which will encounter a large number of CEMP stars, and will greatly speed up the process of determining [Fe/H] as the EWs only need to be measured in two or three lines in relatively clean regions compared to dealing with numerous blended Fe lines. }


   \maketitle

\section{Introduction}

Population III stars are considered to be the first generation of stars. They mostly consist of hydrogen, helium, and negligible amounts of lithium that formed in the Big Bang. As these first generation stars were likely  massive ($\sim$100\,M$_{\rm \odot}$ ;  for discussions on mass see \citealt{Hartwig2018,Clark2011}), their lifespans were short. As such, observational evidence of these first astronomical objects is very difficult to come across and extremely large telescopes are needed to look that far back in time. Knowledge of the first stars must then be indirectly inferred from a second generation of stars. Stellar nucleosynthesis, resulting from the first generation of stars that exploded as supernovae or supernova remnant (neutron star) mergers, contaminate the interstellar medium (ISM) with their produced elements. It is this material from which the second generation of stars emerges and to a great extent it also preserves these elements in their photospheres. 
Excellent study cases of such are the bona fide second generation carbon-enhanced metal-poor (CEMP) stars.

The CEMP stars are enriched in carbon [C/Fe] $\geq$ 0.7 \citep{Beers2005,Aoki2007}.
The relatively high content of carbon in these CEMP stars indicates that carbon is produced early on in the history of the Universe. Hence, the CEMP stars provide us with important constraints on the formation and evolution of the early chemistry in galaxies.

There are four sub-groups of CEMP stars that commonly share two criteria, a carbon abundance of [C/Fe] $\geq$ 0.7 and a metallicity cut of [Fe/H] $\leq -2.0$, which we adopt in this paper. A third criterion involves the abundance of neutron-capture elements, in particular Ba and Eu \citep[or alternatively Sr; ][]{Hansen2019}. 

CEMP$-s$ stars are enriched with $s$low neutron-capture elements and are classified by the abundance [Ba/Fe] $ > 1.0$ \citep{Beers2005,Hansen2016}. The observed abundance pattern is most likely the result of a binary system, where $s-$process rich material is transferred from the envelope of an asymptotic giant branch (AGB) star to a companion star \citep{Lucatello2005,Lugaro2012,TTHansen2016s,Abate2018}. In turn, the companion star becomes a CEMP-s and the AGB star ends its life as an unseen white dwarf while still creating radial velocity variations.

CEMP$-r$ stars are enhanced with $r$apid neutron-capture elements and are defined by [Eu/Fe] $\geq1.0$ \citep{Beers2005,Masseron2010}. As CEMP-r stars are typically single stars \citep{Hansen2015}, it is argued that their chemical composition must stem from the ISM from which it emerges. Furthermore, there must be neutron-rich sites where events, such as rare supernovae type-II explosions or neutron star-neutron star mergers, occur. 

CEMP$-r/s$ (or CEMP$-i$) stars show increased abundances in both r- and s-process elements and therefore an interval is set to [Ba/Eu] = [0.0;0.5] \citep{Beers2005}. \citet{Hampel2016} were able to reproduce the observed abundance patterns of 20 CEMP$-r/s$ stars with an $i$ntermediate neutron-capture process. They modelled the nucleosynthesis of the observed abundance pattern by using fixed neutron densities between 10$^{12}$ \rm $\sim$10$^{15}$ cm$^{-3}$, which may be found during thermal pulses in low-metallicity AGB stars. 

If a CEMP star does not exhibit any significant abundance enhancement of either $s-$ or $r-$process elements, it is denoted as a CEMP-no star and must comply with [Ba/Fe] $< 0.0$. Possible formation sites are faint core-collapse fall-back supernovae \citep{Spite2013,Yong2013,Bonifacio2015} or fast-rotating massive stars \citep{Meynet2010,Frischknecht2016,Choplin2016no}. The CEMP-no are believed to be the true second generation stars.

\begin{figure}
   \centering
    \includegraphics[width=\hsize]{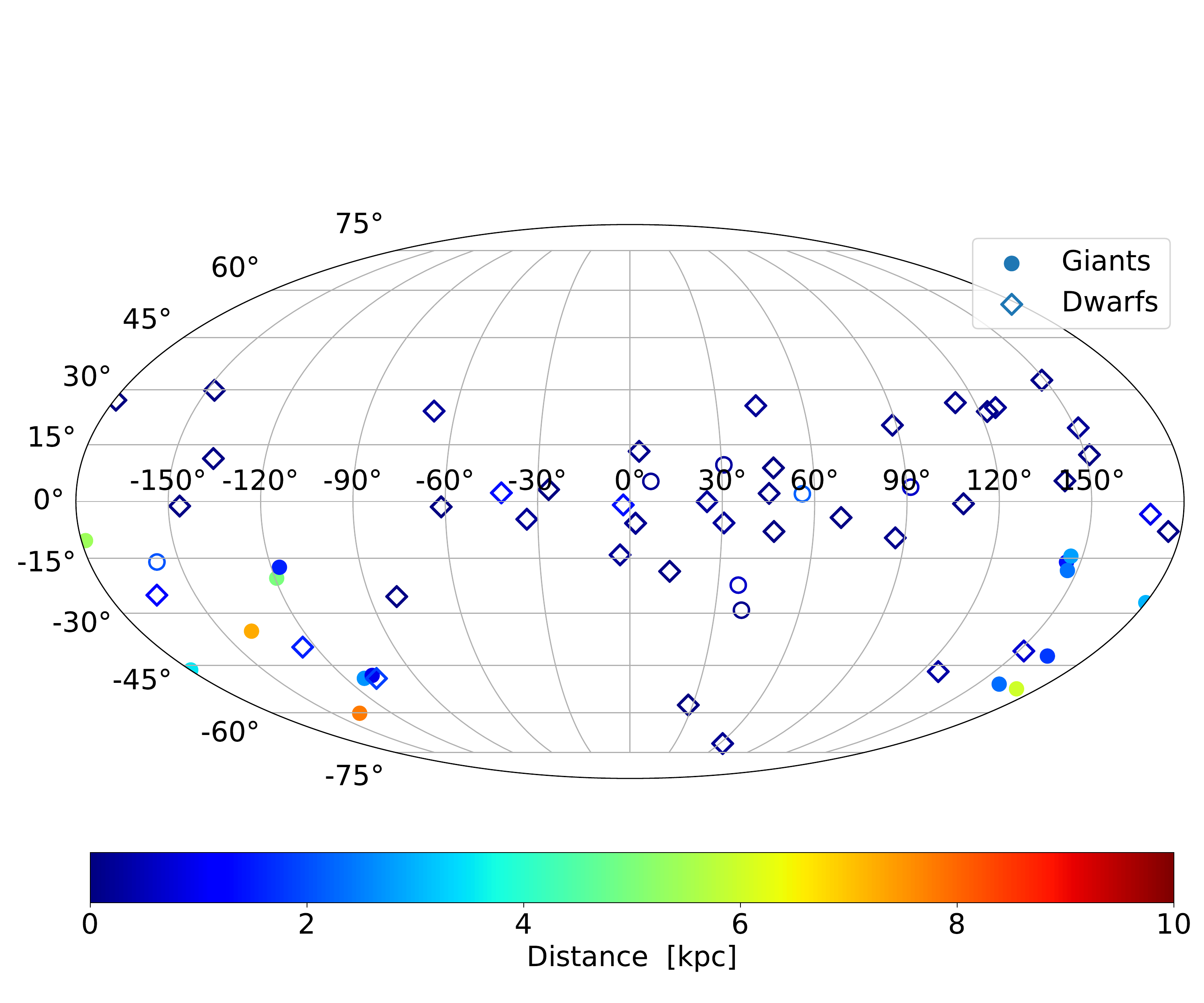}
      \caption{Equatorial coordinates and parallaxes of 64 stars listed in Table \ref{tab:giants} and \ref{tab:dwarfs}, positioning them on the sky and specifying their distances. Stars that have a relative parallax error >\,20\,\% were excluded from this figure. Open symbols indicate high-resolution spectra while closed symbols show low-resolution spectra. Data from the DR2 of the \cite{GaiaDR2}.}
      \label{gaia}
   \end{figure}

Most CEMP stars are faint and located in the Milky Way halo (see Fig.~\ref{gaia}).
Hence, CEMP stars are typically observed when using large telescopes with efficient (possibly low-resolution) spectrographs. 
Further complications in analysing their spectra are introduced by the physics that is inherent to CEMP stars. The low abundances of metals directly correlates with weaker absorption lines. The low temperature that CEMP stars exhibit (see Table \ref{tab-star-para}) allows for molecules to form in the photosphere. These molecules are clearly evident in the spectra as they produce absorption bands spanning broad wavelength ranges of up to $\sim200$ Å, thus effectively rendering parts of the spectra unusable for accurate atomic abundance analysis (see Fig.~\ref{molecularbands}).
The low metallicity, cool temperatures in combination with the low resolution make these spectra prone to the blending of their absorption lines and molecular features, which makes measuring the weak Fe lines, and in turn determining [Fe/H] in CEMP stars, extremely challenging. 
Previous studies have shown that low-resolution spectrum analyses may result in abundances that are 0.3$-$0.4\,dex higher compared to the high-resolution studies of the same stars \citep[e.g.][]{Hansen2019,Aguado2016,Caffau2011}. 

In order to understand CEMP stars and their origin, we must make use of chemical analysis. In doing so, we generally utilised 1D, local theormodynamical equilibrium (LTE) model atmospheres. These model atmospheres require the following four key parameters: effective temperature (T), surface gravity (logg), metallicity ([Fe/H]), and microturbulence ($\xi$). These parameters are interdependent and require that their values be determined through iterative processes. The temperatures can be determined from photometry, with the possible exception of CEMP stars, and the gravity from parallaxes, such as $\omega$ or isochrones.\ Whereas, metallicity typically relies on the spectra if we neglect the use of Str\"omgren photometry, for instance. The [Fe/H] is determined by measuring the equivalent widths (EWs) of Fe absorption lines in spectra. This step is increasingly difficult when the uncertainties it introduces propagate undesirably, thus affecting the other parameters and ultimately any abundance results that are derived. 
Finally, the microturbulence is typically set by balancing the trend of Fe-abundances with EW or by the use of empirical scaling relations \citep[e.g.][]{Mashonkina2017a}.

In large samples, the classical methods, relying solely on spectra or a mixture of spectra and photometry, become much too slow and automated codes are needed to speed up the process. However, most of the currently existing codes fail when dealing with CEMP stars, owing to the excessive line blends from the molecular bands \citep[e.g.][]{Hanke2018}. All such codes need large training samples with accurate stellar parameters.
To improve the [Fe/H] estimates in CEMP stars, we have therefore developed an empirical method that yields [Fe/H] values more accurately than by determining the value from EWs of (blended) Fe lines. Moreover, to somewhat circumvent the dependence of metallicity on the remaining parameters, we used the Infrared Flux Method (IRFM) and selected the least [Fe/H] sensitive colour in order to obtain the effective temperature, isochrones that we then used to determine the surface gravities, and the microturbulence that was calculated using an empirical relation.

We follow the approach of \cite{Wallerstein2012}, who
from the Gaia data, anticipated thousands of observations of RR Lyrae stars with a narrow wavelength range at a medium resolution of $\sim$11500. They developed a method that correlates the EW of the Ca II line at 8498 Å and the metallicity of RR Lyrae stars. Drawing from the same principles, we investigate the relation between a CEMP star's metallicity and another prominent elemental spectral feature, thereby significantly reducing the time required to determine metallicities. This will highly benefit future surveys in that training samples are better and larger, and one can simply rely on using the same spectral features as suggested here. We also show that the relations can be used for C-normal stars. In both cases, the relations are valid in the span from $-3.2<$[Fe/H]$\leq-0.0$ (where the exact range is trace element dependent).

The paper is structured as follows. Section~\ref{sec:obs} describes our sample stars in groups, detailing the observations and data treatment.
Section~\ref{sec:analysis} summarises the analysis, the determination of stellar parameters, and their uncertainties. Section~\ref{sec:results} presents our new metallicity tracer for CEMP and C-normal stars and shows the results of our relations. Finally, Sect.~\ref{sec:discussion} discusses and concludes on the validity and applications of the relations.

\section{Observation}\label{sec:obs}

All stars in our samples were observed with the Very Large Telescope (VLT) at Paranal, Chile, using either the UVES or X-SHOOTER echelle spectrographs. A schematic overview of the sample is shown in Fig.~\ref{gaia}.

In Table \ref{tab-star-para}, our sub-sample consisting of 28 poorly studied, mainly C-rich stars is listed. 
We also include two well-studied dwarf stars as these have been suggested to be CEMP stars as well \citep{Placco2016}. 
The two well-studied metal-poor dwarfs are G64-12 and G64-37. They are extremely metal-poor and there is an ongoing debate as to whether or not these stars are carbon-enhanced stars \citep{Placco2016,Amarsi2019,Norris2019}. We include these stars in our sample as reference points in terms of stellar parameters and in order to settle the CEMP dispute.
These stars were observed on 17 June 2006 and 15 April 2000, respectively, using VLT/UVES. A slit width of 1" \hspace{0.2mm} in UVES yields a resolving power of $\sim$47000 in the ultraviolet (UV), and an exposure time of 2877 and 2700\,s results in a signal-to-noise ratio (S/N) of $\sim$180 and $\sim$220 per pixel at 5000 Å, respectively. The raw echelle data of both EMP stars were reduced with the UVES pipeline. It automatically performs data reduction in terms of extracting a 1D spectrum and merging of orders, containing both a UV and visual part of the spectrum. Our data reduction then required wavelength shifting, co-adding multiple exposures, and normalising the continuum. 

\begin{figure}
   \centering
   \includegraphics[width=\hsize]{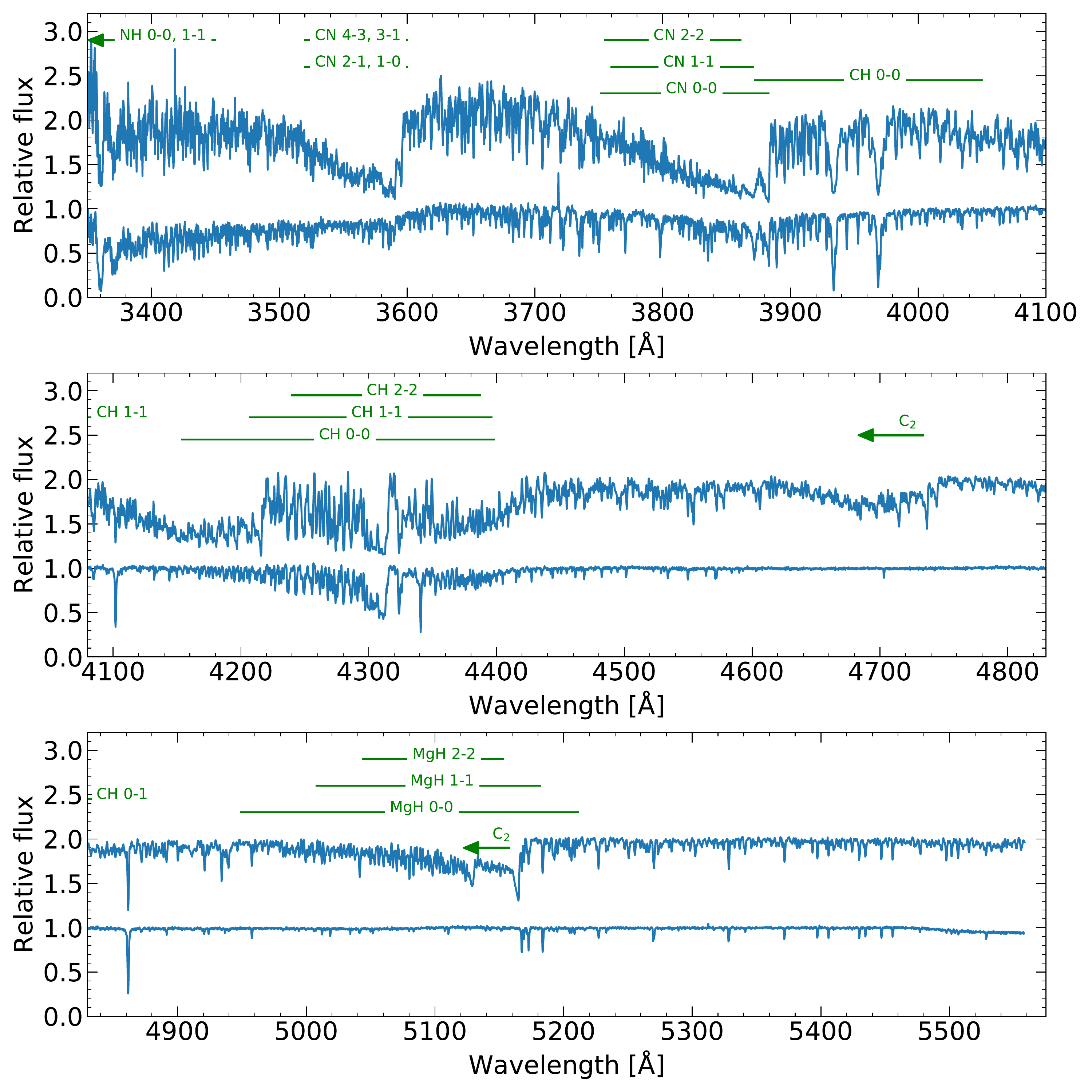}
      \caption{Spectral features in the X-Shooter spectra of the CEMP$-s$ star \object{HE0253-6024} (offset flux level) and the extremely metal-poor CEMP-no star \object{HE0020-1741}.}
         \label{molecularbands}
   \end{figure}

The 28 less studied giant stars can be categorised into two sub-groups. The first sub-group was observed between 24 March 2010 and 17 August 2010 and has lower signal to noise, and the second sub-group was observed between 14 October 2012 and 26 January 2013 (with higher S/N, \citealt{Hansen2019}). Both sub-groups were observed with the X-SHOOTER spectrograph with a slit width of 1", \hspace{0.2mm} which results in a resolving power of $\sim$5400 in the UV. The signal to noise of the stars from these sub-groups ranges from $\sim$6 to $<200$ at 5000 Å. In terms of data reduction, we similarly shifted the wavelength to the rest frame and normalised the continuum.
With the low signal to noise, low resolution, low temperatures, low metallicity, and molecular bands, we see in Fig.~\ref{molecularbands} that  the wavelength ranges, in which we are able to measure Fe lines, are very sparse.

In addition we included a third group of stars, mostly C-normal, consisting of 42 dwarfs and ten giants observed at high resolution. Sixteen of the dwarfs were observed in 2015 using UVES with a slit width of 0.8",\hspace{0.2mm} providing a resolution of $\sim$50000 in the UV, with a signal to noise of $\sim$130 and 250 at 4000\,Å and 6000\,Å, respectively. All spectra were fully reduced \citep{Reggiani2017}. We focused on a small wavelength range (4290 - 4350\,Å) to exclusively measure carbon and derive an abundance from the CH molecular band. The spectra of the last 26 dwarfs are taken from the ESO archive and reduced (Reggiani et al. in prep).
 
For all 42 dwarfs, we adopted the stellar parameters from \cite{Reggiani2017} and Reggiani et al. in prep, see Table~\ref{tab:dwarfs}. 
Finally, we included ten well-studied giants for comparison and calibration purposes. The spectra are also taken from the reduced ESO archive and \cite{Hansen2012} and the stellar parameters are also adopted from the literature (Table~\ref{tab:giants} and \ref{tab:dwarfs}).

Hence, the total sample of 82 stars consists of well-known stars observed at high resolution and poorly-known stars observed at low and medium resolution, encompassing a broad range of C abundances. 
Fig. \ref{gaia} shows the positions and distances of 64 sample stars. The distances were calculated from the inverse parallaxes retrieved from the Gaia Data Release 2 \cite{GaiaDR2}, stars with a relative parallax error $\left(\sigma_{rel. \hspace{1mm} error} = \omega_{error}/\omega \right)$ of more than 20\% were removed.

\begin{table}

\caption{Temperature, metallicity, surface gravity, microturbulence, and classification. Parentheses indicate less certain classification. The '*' shows a well-classified star used for uncertainty computations and shown in Fig. \ref{molecularbands}.}
\label{tab-star-para}
\centering
\resizebox{\hsize}{!}{\begin{tabular}{c c c c c c}

\hline \hline
\noalign{\smallskip}
Star    &       T$_{\text{eff}}$        &       [Fe/H]  &       log \textit{g}         &        $\xi$  &   Classification      \\ 
        &       [K]     &       [dex]   &       [cgs]   &       [km s$^{-1}$]   &     \\ 
\noalign{\smallskip}
\hline
\noalign{\smallskip}
HE0002-1037     &       4959    &       -2.47   &       1.93    &       1.81    &   CEMP$-r/s$       \\
HE0020-1741     &       4738    &       -3.30   &       1.26    &       1.99    &   CEMP-no  \\
HE0039-2635     &       4970    &       -3.20   &       1.90    &       1.80    &   CEMP$-s$ \\
HE0059-6540     &       4983    &       -2.15   &       1.97    &       1.77    &   CEMP$-r/s$       \\
HE0221-3218     &       4760    &       -0.80   &       2.50    &       1.60    &   MP       \\
HE0241-3512 &   4607    &       -1.76   &       1.11    &       2.02    &   MP/(CEMP$-s$) \\
\hspace{0.8mm} HE0253-6024*     &       4640    &       -2.00   &       1.17    &       2.01    &   CEMP$-s$ \\
HE0317-4705     &       4692    &       -2.25   &       1.23    &       2.02    &   (CEMP$-r/s$)     \\
HE0400-2030 &   5160    &       -2.10   &       2.45    &       1.64    &   CEMP$-s$      \\
HE0408-1733 &   4525    &       -0.75   &       1.11    &       1.83    &   MP    \\
HE0414-0343 &   4633    &       -2.30   &       1.10    &       2.07    &   CEMP$-s$   \\
HE0430-4901     &       5500    &       -3.10   &       3.30    &       1.50    &   CEMP$-s$ \\
HE0448-4806 &   5750    &       -2.22   &       3.54    &       1.47    &   CEMP$-s$      \\
HE0516-2515 &   4400    &   -2.50    &   0.70    &   2.20    &   (CEMP$-s/$no) \\
HE1431-0245 &   5200    &       -2.50   &       2.30    &       1.80    &   CEMP$-s$      \\
HE2138-1616 &   4900    &       -0.50   &       1.90    &       1.60    &   MP    \\
HE2141-1441 &   4576    &       -0.38   &       1.24    &       1.67    &   MP    \\
HE2144-1832 &   4200    &       -1.70   &       0.60    &       2.20    &   MP/(CEMP$-s$) \\
HE2153-2323 &   4300    &       -2.46   &       0.60    &       2.30    &   CEMP$-s$      \\
HE2158-5134     &       4915    &       -2.70   &       1.77    &       1.85    &   CEMP$-s$ \\
HE2235-5058     &       5110    &       -2.70   &       2.32    &       1.68    &   CEMP$-s$ \\
HE2250-4229     &       5101    &       -2.70   &       2.28    &       1.71    &   VMP/(CEMP-no)    \\
HE2258-4427     &       4560    &       -2.10   &       1.00    &       2.10    &   CEMP$-s$ \\
HE2310-4523     &       4695    &       -2.40   &       1.24    &       2.02    &   VMP      \\
HE2319-5228     &       4817    &       -2.80   &       1.49    &       1.93    &   CEMP-no  \\
HE2339-4240     &       5046    &       -2.33   &       2.13    &       1.75    &   CEMP$-s$ \\
HE2357-2718     &       4490    &       -0.66   &       1.08    &       1.81    &   MP       \\
HE2358-4640     &       5111    &       -1.75   &       2.34    &       1.65    &   MP       \\
\noalign{\smallskip}
\hline
\noalign{\smallskip}
G64-12  &       6400    &       -3.30   &       4.15    &       2.0    &   EMP   \\
G64-37  &       6550    &       -3.10   &       2.25    &       1.6    &   EMP   \\
\noalign{\smallskip}
\hline
\end{tabular}}
\end{table}

\section{Analysis}\label{sec:analysis}
\subsection{Data reduction software}

The data reduction was completed using our own software in Python, which performs the following steps and functionalities:

First, reads in and plots both 1D and 2D spectra.

Second, doppler shifts the spectrum using either the magnesium triplet at $\sim$ 5183 Å  or the calcium triplet $\sim$ 8542 Å, depending on spectral coverage.

Third, normalises a part or the full range of a spectrum by fitting a pseudo continuum to the data points by using a chebyshev polynomial and then by dividing the spectra with this pseudo continuum. Options include weights of user specified wavelength intervals and forcing the fitting to only be done on continuum and feature-free regions (see Fig.~\ref{molecularbands}), effectively fitting the continuum better and bypassing spectral absorption features. Notably, we use the following regions: 4450 - 4550, $\sim5220 - 5480$, and 5650 - 6200\,\AA\ (and 4750 - 4800, 6200 - 6900, and 7700 - 7840\,\AA\ as they are more or less free of molecular bands, but thy may contain tellurics).

Fourth, measures EWs by fitting a Gaussian profile to the spectral feature in question, and

Fifth, the co-adding of spectra was done by calculating the mean of the added flux values. 

We tailored specific features to automate several processes in order to analyse a large number of CEMP spectra in a time efficient manner. Identifying lines is done by using the NIST\footnote{https://www.nist.gov/pml/atomic-spectra-database} database for visual recognition of specific elements plotted directly onto the spectra. The solar and Arcturus spectra are plotted against all other spectra for visual inspection. Finally, determining the continuum is vastly improved by selecting band free regions and is done in four different ways using 25, 50, 100, and 200\,\AA\ ranges.

\subsection{Stellar parameters}\label{stelpar}
\begin{figure}
   \centering
   \includegraphics[width=\hsize]{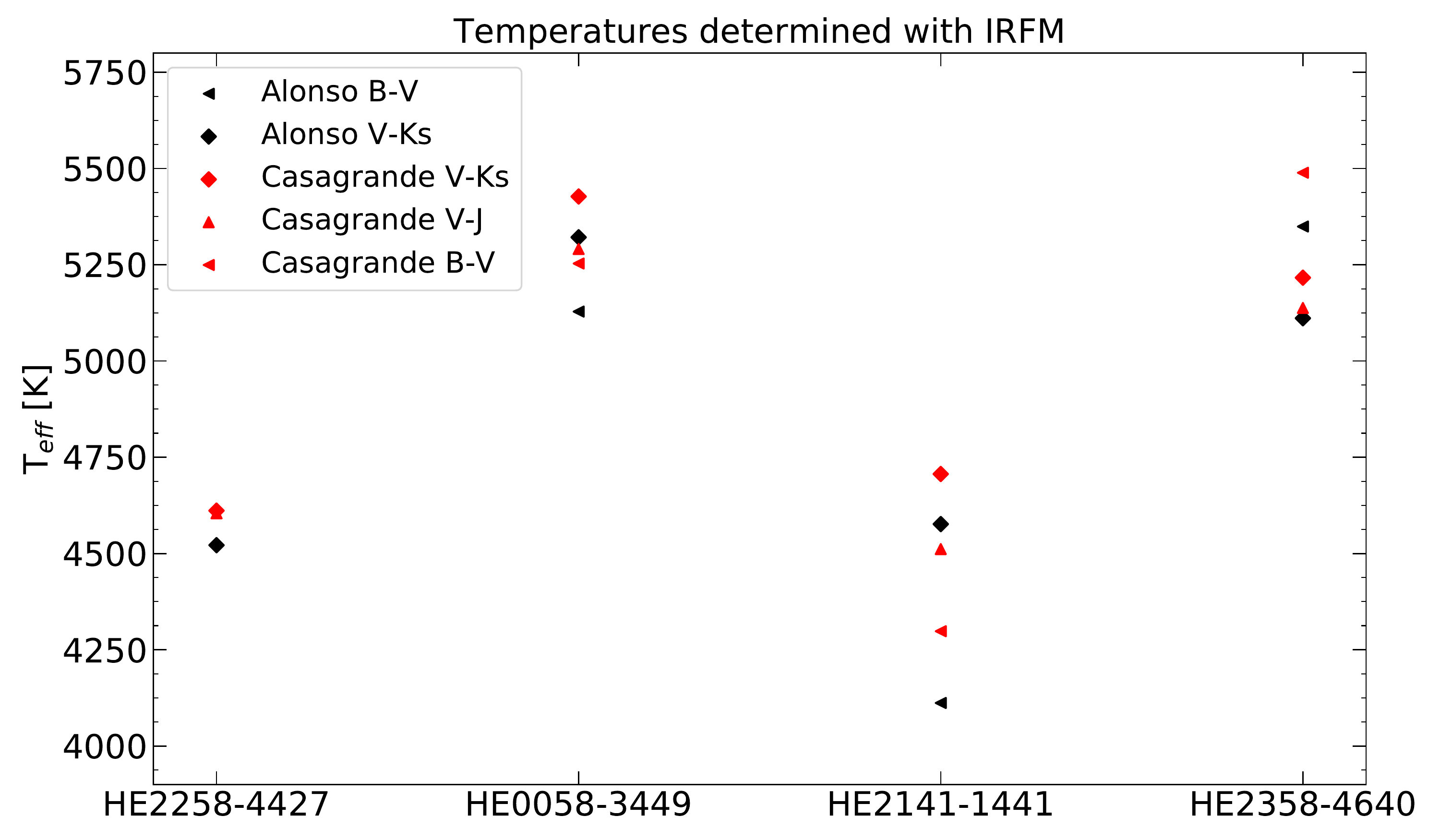}
      \caption{Comparison of temperatures derived from different IRFM relations - see legend for details.}
         \label{red_dered}
   \end{figure}

The four governing parameters used in 1D LTE models are temperature, surface gravity, metallicity, and microturbulence. Temperatures were calculated by using the IRFM, adopting the photometric colour $V - K_{s}$ as this is fairly unaffected by [Fe/H] \citep{Alonso1999}, and by employing the mean extinction by \cite{Schlafly2011} from the Infrared Science Archive (IRSA\footnote{https://irsa.ipac.caltech.edu/applications/DUST/}). Since the mean extinction was given in $E(B-V),$ we converted it to $E(V-K_s)$ with the extinction law provided by \cite{Alonso1996} and \cite{Bessel2005}.  For the colour-temperature-metallicity relation, we investigated whether or not the relation from \cite{Casagrande2010} would be applicable. We found their relation to systematically be 100 K higher than the temperature from \citet{Alonso1999} (adopted) for the same colour, see Fig.~\ref{red_dered}.
 
\begin{figure}
   \centering
    \includegraphics[width=\hsize]{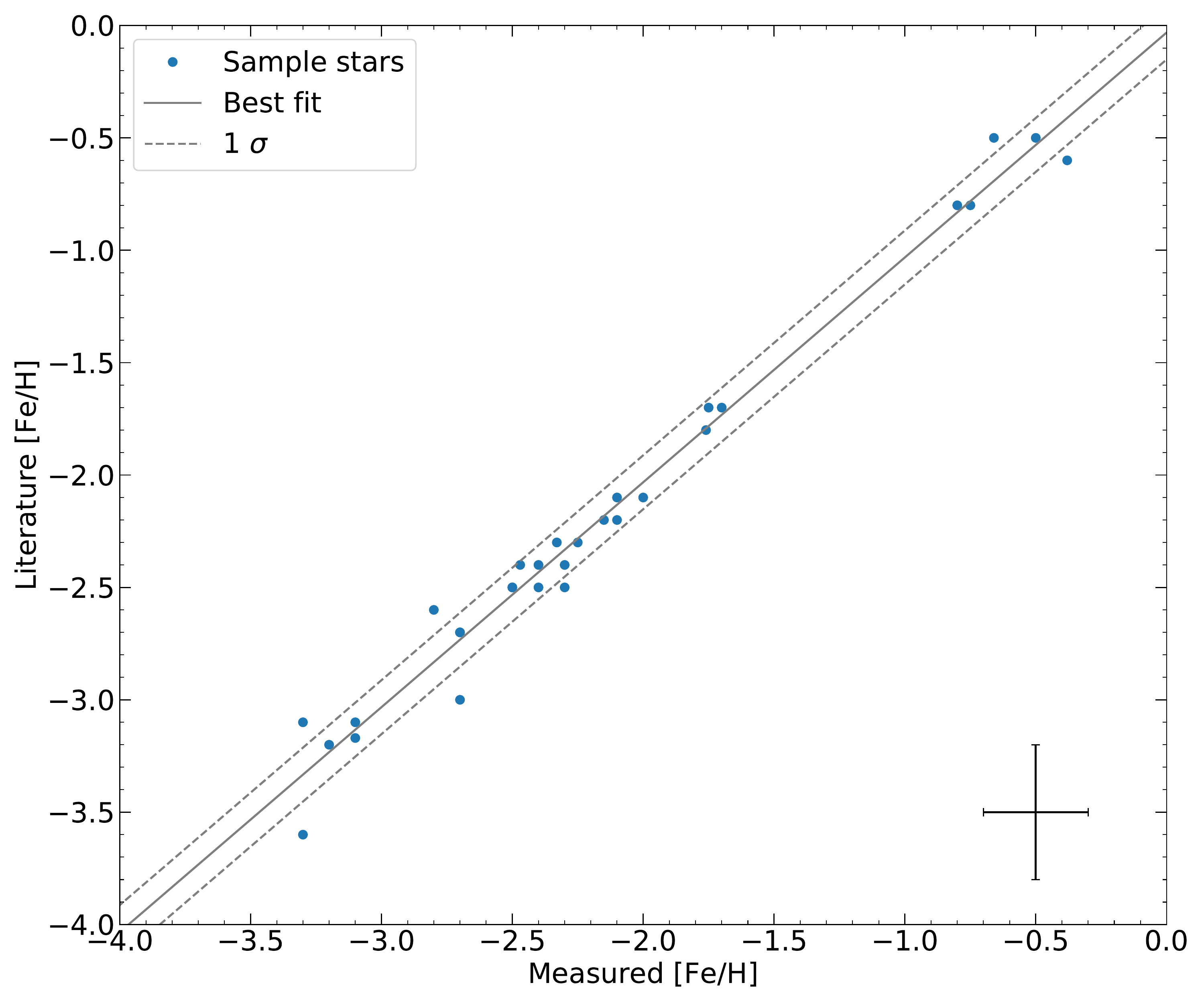}
      \caption{Our [Fe/H], of all 28 stars from Table \ref{tab-star-para}, based on EWs of Fe lines compared to \citet{Hansen2016,Hansen2019}. A clear 1:1 linear correlation is seen with a low scatter ($<\sigma$). A general uncertainty is plotted in the lower right corner.}
         \label{fehfeh}
   \end{figure}

Surface gravities were obtained when fitting isochrones (D. Yong priv. comm.). In order to derive the microturbulence, we used an empirical relation (M. Bergemann priv. comm.), which relates the temperature, surface gravity, and metallicity to the microturbulence, where we used our initial guess for the metallicity to obtain an estimate for the temperature, surface gravity, and microturbulence. The metallicities are determined by measuring the EWs of selected (cleaner) Fe I lines (see Table \ref{tab:Felines}) and thus this becomes an iterative process until the final values for the parameters are obtained.
\begin{table}
 \caption{Wavelength, excitation potential, and oscillator strength for Fe I lines used for deriving [Fe/H].}
 \label{tab:Felines}
 \centering
\begin{tabular}{c c c c}
\hline \hline
\noalign{\smallskip}
Wavelength      &       Species &       EP      &        log\textit{gf} \\ 
$[$\AA\,$]$     &       &       $[$eV$]$        &       $[$dex$]$       \\ 
\noalign{\smallskip}
\hline
\noalign{\smallskip}
4920.503        &       26.0    &       2.833   &       +0.068  \\
4973.102        &       26.0    &       3.960   &       $-0.920$        \\
5208.594        &       26.0    &       3.241   &       $-0.897$        \\
5227.189        &       26.0    &       1.557   &       $-1.228$        \\
5232.940        &       26.0    &       2.939   &       $-0.057$        \\
5307.361        &       26.0    &       1.608   &       $-2.987$        \\
5322.040        &       26.0    &       2.279   &       $-2.803$        \\
5341.024        &       26.0    &       1.607   &       $-1.953$        \\
5371.489        &       26.0    &       0.958   &       $-1.645$        \\
5383.369        &       26.0    &       4.312   &       +0.645  \\
5393.167        &       26.0    &       3.241   &       $-0.715$        \\
5397.128        &       26.0    &       0.914   &       $-1.993$        \\
5405.775        &       26.0    &       0.990   &       $-1.844$        \\
5434.524        &       26.0    &       1.011   &       $-2.122$        \\
6027.051        &       26.0    &       4.076   &       $-1.089$        \\
8824.220    &   26.0    &       2.198   &       $-1.540$        \\
\noalign{\smallskip}
\hline
\end{tabular}
\end{table}
To quantify the accuracy of our  derived metallicities from a conventional method (i.e. measuring EWs of Fe lines), we compare our metallicities to literature \citep[][see Fig.~\ref{fehfeh}]{Hansen2016,Hansen2019}. 

In Fig.~\ref{fehfeh} we find the residual standard deviation (RSD) to be 0.12 dex, which is reasonable given the difference in data reduction and spectrum analysis. From this reasoning, we also quantify how accurate our derived metallicities are. From the standard deviations (SD) of the metallicities calculated for each star, which are based on the number and abundances of the Fe lines, our mean of all the sample stars' standard deviations 
is 0.27 dex when using the EWs of Fe lines. It is, therefore, unlikely to determine metallicities of CEMP stars to a better degree than $\pm$ 0.27\,dex (in low S/N spectra, see, e.g. \citealt{Hansen2016}).

\subsection{Error propagation}

We conduct a detailed analysis of the uncertainties on our derived stellar parameters and abundances for \object{HE0253-6024}. The uncertainty on temperature from IRFM methods are driven by uncertainties on the observed colours and de-reddening. These amount to 43\,K. The method has an internal uncertainty  and since we do not know the exact impact of strong bands on the IRFM method, we adopt a larger, conservative value of 100\,K (see Table~\ref{tab-star-para2}). We explore the impact the  temperature uncertainty has on the surface gravity. A 100\,K change corresponds to a change in logg of 0.22 dex. We use this as an uncertainty since the metallicity does not significantly add to this uncertainty. The uncertainty on the microturbulence is derived by changing the input values by these uncertainties. This amounts to $\sim \pm 0.1$\,km/s. We use the statistical errors on [Fe/H] from the averaged standard deviation of the line-to-line spread ($\pm 0.27$\,dex), which indirectly includes uncertainties in atomic data (e.g. excitation potential and the oscillator strength), continuum placement, and the EW fitting of the Fe lines. The impact of these uncertainties is listed in Table~\ref{tab-star-para2}.

We can also estimate the uncertainties on the measured EWs at all S/N of our sample stars (6 -- 200), using Cayrel's formula \citep{Cayrel1988,Cayrel2004}
It becomes obvious that at the lowest S/N, the uncertainties in EW and in turn [Fe/H] are driven by noise; whereas above S/N$\sim25,$ the atomic data and  continuum placement causing line-to-line scatter dominate the uncertainties.

Besides the four stellar parameters, the continuum placement has a considerable impact on the derived abundances. It can affect the derived abundances by as much as 0.1 dex. Since the stars in question are rich in carbon, we investigated whether the amount of carbon influences the derived Ba and Eu abundance. This is not the case since we used the red Ba and Eu lines at 5853\AA\, and 6645\AA,\, which are located in regions that are fairly free of molecular bands (see Table \ref{tab-star-para2}).

\begin{table*}
 \caption{Uncertainties on stellar parameters and abundances of the star HE0253-6024.}
 \label{tab-star-para2}
 \centering
\begin{tabular}{c c c c c c c c c}
\hline \hline \noalign{\smallskip}
        &               &       $\Delta$ T      &       $\Delta$ log \textit{g}  &       $\Delta \xi$    &       $\Delta$ [Fe/H]         &       $\Delta$ [C/Fe]  &       $\Delta$ [Ba/Fe] &      $\Delta$ [Eu/Fe] \\
        &                       &       [K]     &       [cgs]   &       [km s$^{-1}$] &     [dex] & [dex]   &       [dex]   &       [dex]           \\ \noalign{\smallskip} \hline \noalign{\smallskip}
T       &       $\pm$ 100 K     &       …     &       0.22    &       0.01    &       0.18    &       0.17    &       0.08    &       0.06    \\
$\left[ \text{Fe/H} \right] $   &       $\pm$ 0.27 dex  &       4       &       0.03    &       0.03    &       …     &       0.01    &       0.01    &       0.02    \\
logg    &       $\pm$ 0.22 dex  &       …     &       …     &       0.08    &       0.06    &       0.02    &       0.05    &       0.07    \\
$\xi$   &       $\pm$ 0.1 km/s  &       …     &       …     &       …     &       0.05    &       0.01    &       0.02    &       0.02    \\
EP      &       $\pm$ 0.05 eV   &       …     &       …     &       …     &       0.06    &       …     &       …     &       …     \\
Log gf  &       $\pm$ 0.025     &       …     &       …     &       …     &       0.03    &       …     &       …     &       …     \\
EW      &       $\pm$ 10 mÅ    &       …     &       …     &       …     &       0.11    &       …     &       …     &       …     \\
Continuum       &               &       …     &       …     &       …     &       …     &       0.03    &       0.10    &       0.10    \\
$[\text{C}/\text{Fe}]$          &       $\pm$ 0.3 dex   &       …     &       …     &       …     &       …     &       …     &       0.03    &       0.0     \\
E(B-V)  &       $\pm$ 0.01 mag  &       30      &       …     &       …     &       …     &       …     &       …     &       …     \\
(V-K)   &       $\pm$ 0.03 mag  &       30      &       …     &       …     &       …     &       …     &       …     &       …     \\ \noalign{\smallskip} \hline \noalign{\smallskip}
Total   &               &       43 (100)        &       0.22    &       0.09    &       0.23    &       0.26    &       0.21    &       0.14    \\ \noalign{\smallskip} \hline

\end{tabular}
\end{table*}

 \subsection{Reference and CEMP stars}
Our sample includes a mixture of C-rich and C-normal stars so as to probe if these two, differently, chemically enriched groups of stars share the same metallicity tracer. We therefore included, amongst others, \object{G64-12} ([Fe/H]=$-3.3$) and \object{G64-37} ([Fe/H]=$-3.1$). These two stars are well-studied stars (and part of the Gaia-ESO benchmark stars), yet their chemical composition is still disputed to some extent. \cite{Placco2016} claim that these two stars are CEMP stars based on a 1D, LTE spectrum synthesis of the CH G-band, while \cite{Amarsi2019} counters with a 3D, non-LTE (NLTE) analysis of the red atomic lines ($>9000$\,\AA), yielding a [C/Fe]$<0.2$ for both stars. 

We have re-analysed the high-resolution, high S/N spectra of \cite{Hansen2012} and the even higher quality spectra on which \cite{Placco2016} base their study. We find [C/Fe] = 0.50 for \object{G64-12} from the CH G band, clearly showing that this star behaves like C-normal extremely metal-poor halo stars. We note that the CH features are so weak (almost around the noise level), that a change in flux of +0.1\% results in +0.1\,dex in C abundance. Hence, the continuum placement becomes extremely important. The lines are similarly weak for \object{G64-37}. Here, a very slight slope in the normalised continuum prevents fitting the band head. Features bluewards of the head yield very uncertain values of [C/Fe] $\sim$ 0.6, while features redwards seem to indicate 0.1\,dex higher values. Despite the high quality UVES spectra, the spectrum fitting and results are not conclusive and we can at best set [C/Fe]$<0.7$ for this star. Based on this, we would not classify G64-37 as a CEMP star, but we would abstain from a clear classification. We assign the C-enhancement derived in \cite{Placco2016} to noise and slight upturns ($\sim0.003$ in normalised flux) in the spectra, which may be traces of the blaze function that could have left imprints in the spectrum that then affected the continuum placement (around 4280-4300\,\AA\,), rendering it too high in their study.
We do, however, note that when considering the large 3D, NLTE corrections ($\sim 1$\,dex) versus the 1D, LTE abundances shown in \citet{Norris2019}, the values are in agreement with the 3D NLTE abundances from \citet{Amarsi2019} and also with the 1D, LTE study by \citet{Placco2016}. If such corrections would be applied to our values, the two stars would end up having subsolar [C/Fe]. In any case, the enhanced C abundances are an outcome of the simplified 1D, LTE assumptions \citep[see also][for 3D effects on the G-band]{Gallagher2016}.

\begin{figure}
   \centering
    \includegraphics[width=\hsize]{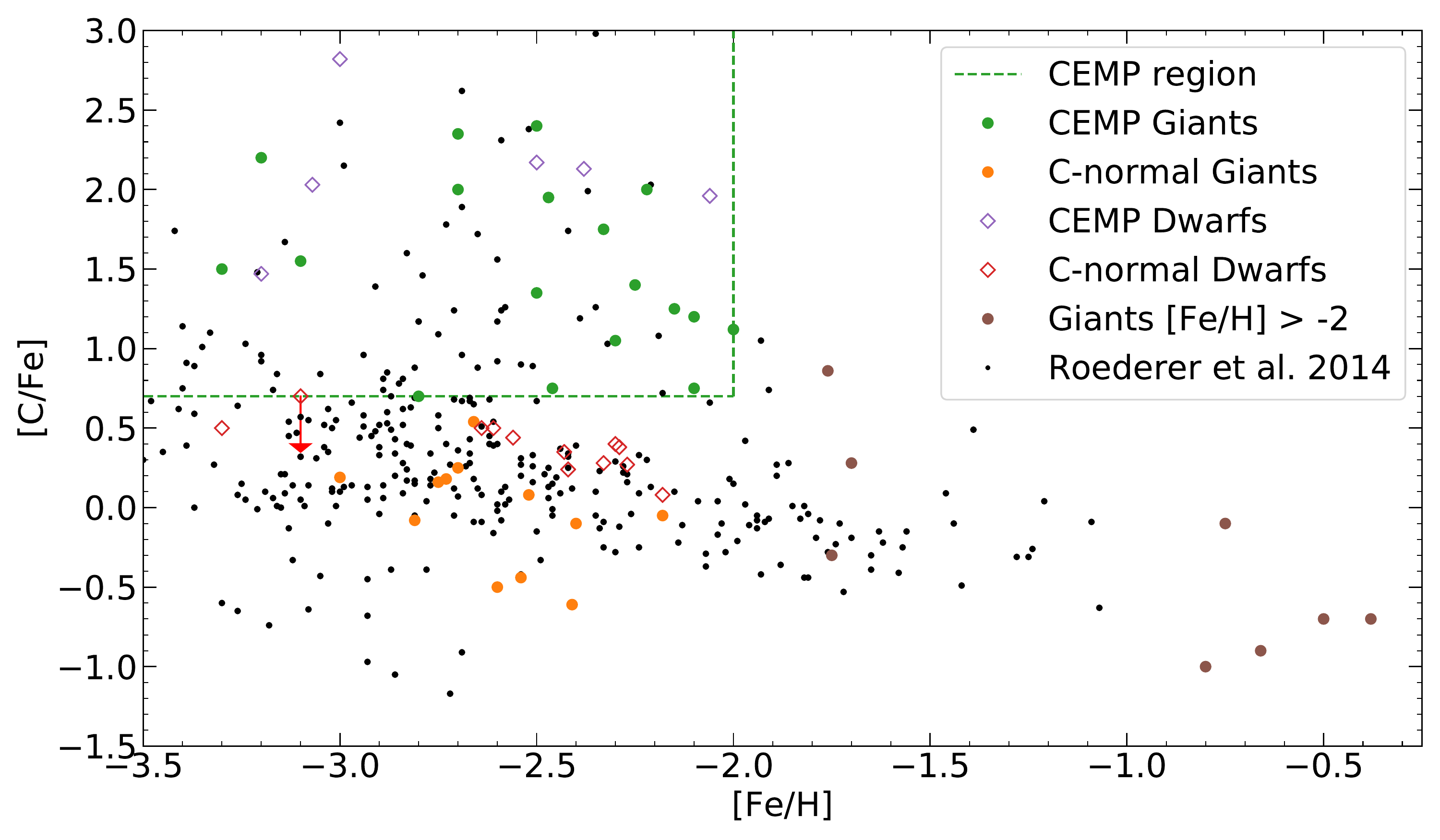}
      \caption{[C/Fe] vs. [Fe/H] for 38 sample stars from Table~\ref{tab:giants} and 18 from \ref{tab:dwarfs}.}

         \label{cfe}
   \end{figure}

In Fig.~\ref{cfe}, 56  stars (sample plus comparison) are plotted against \cite{Roederer2014}, showing their [C/Fe] as a function of [Fe/H]. The box separates our CEMP stars from the C-normal stars. In total we have 38 giants (18 CEMP, 12 C-normal, eight stars above [Fe/H]=$-2$) and 18 dwarfs (six CEMP, ten C-normal, two reference stars). The ten C-normal dwarf stars are from \cite{Reggiani2017}; we derived new C abundances for these stars, finding them to be C-normal (see open symbols in Fig.~\ref{cfe} and Table~\ref{tab:dwarfs}).

\subsection{An empirical metallicity tracer}
\begin{figure}
   \centering
    \includegraphics[width=\hsize]{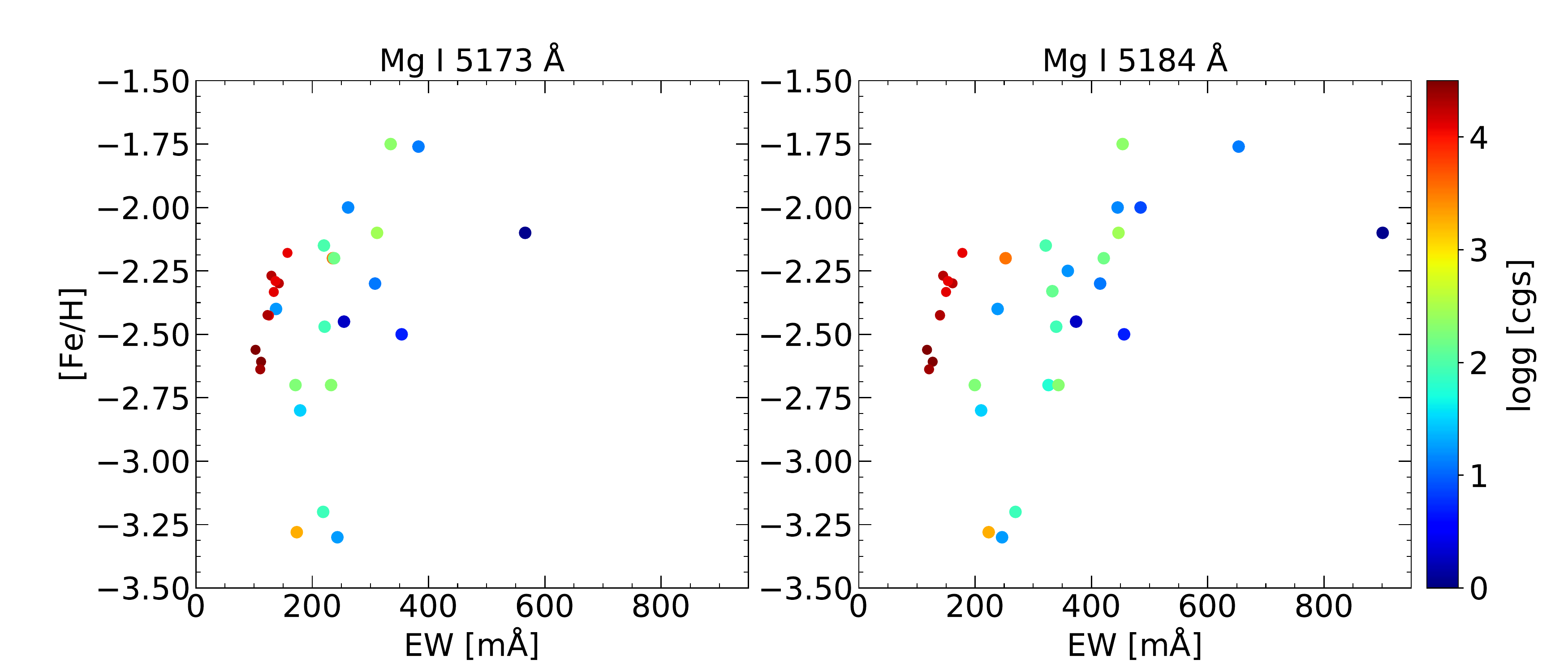}
      \caption{Metallicities vs. equivalent widths with a colourbar, indicating surface gravities for Mg I line 5173 \AA \ and 5184\,\AA\, of the same sample stars (19/23 respectively) as depicted in Fig. \ref{mg_dobbelt}, including ten C-normal dwarfs from Table \ref{tab:dwarfs}.} 
         \label{mgfedobbel}
   \end{figure}

In order to develop a metallicity tracer for CEMP stars, it is necessary to distinguish between dwarfs and giants. Spectral absorption lines exhibit different behaviour depending on the star's surface gravity. 
 One method to distinguish between dwarfs and giants is by measuring the EWs of the Mg I b triplet since the wings are sensitive to surface gravity. The Mg I b triplet is useful as this spectral feature is relatively strong and clearly visible in almost all spectra regardless of S/N, resolution, and down to extremely low metallicities. In CEMP stars, the two redder Mg I lines are available for this analysis as the bluest feature in the triplet is blended with a molecular C$_2$ band. By plotting the EWs of Mg I at 5173 Å and Mg I at 5184 Å against metallicity, we see in Fig.~\ref{mgfedobbel} that dwarfs and giants do indeed group separately, but there is considerable scatter around the line(s)
 from both dwarfs and giants. Any division formulated would either include a subset of stars from both dwarfs and giants or exclude stars from a particular group, ultimately disqualifying this method for distinguishing giants from dwarfs. 
  The best solution is to develop two separate metallicity tracers for CEMP dwarfs and CEMP giants, respectively.

\begin{figure}
   \centering
    \includegraphics[width=\hsize]{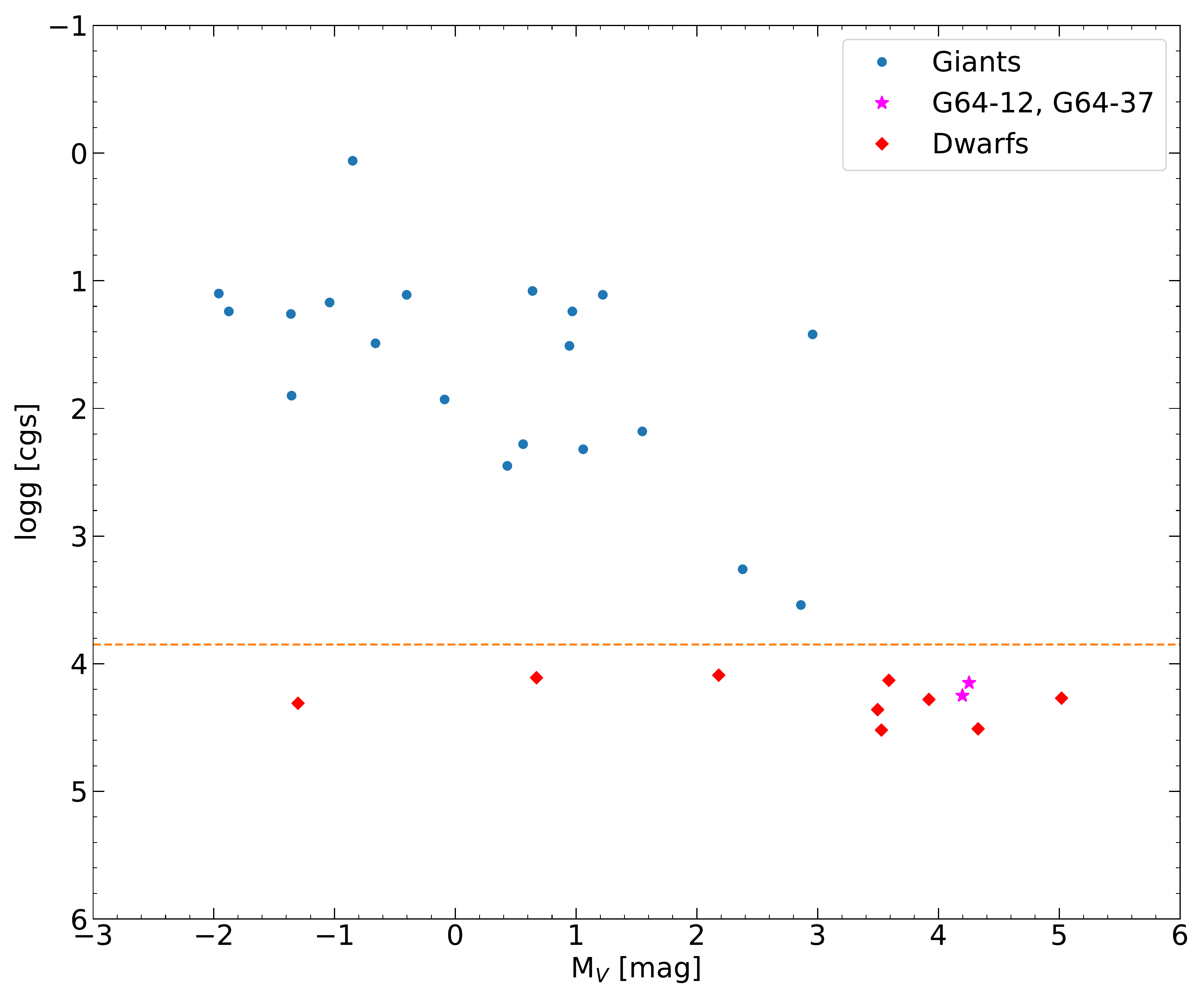}
      \caption{Surface gravity, logg, and de-reddened absolute V magnitude (M$_{V,0}$) for 20 sample stars from Table \ref{tab:giants} and 11 from \ref{tab:dwarfs}.  Data from the DR2 of the \cite{GaiaDR2}. }
         \label{GaiaMv}
   \end{figure}

Another procedure is to use the apparent magnitudes and parallaxes of the stars to calculate their absolute magnitudes. This has become possible because of the Gaia Data Release 2 (April 2018), which contains parallaxes of about 1.7 billion stars \citep{GaiaDR2}. By simply using the parallaxes and small angle approximation, we calculated the distances and thereby the stars' absolute magnitudes. As seen in Fig.~\ref{GaiaMv}, the dwarfs are typically found nearby due to their fainter nature, while the giants are brighter and can be observed further away. We have loosely placed a division (yellow line) between dwarfs and giants in this figure.

\section{Results}\label{sec:results}

  \begin{figure}
   \centering
    \includegraphics[width=\hsize]{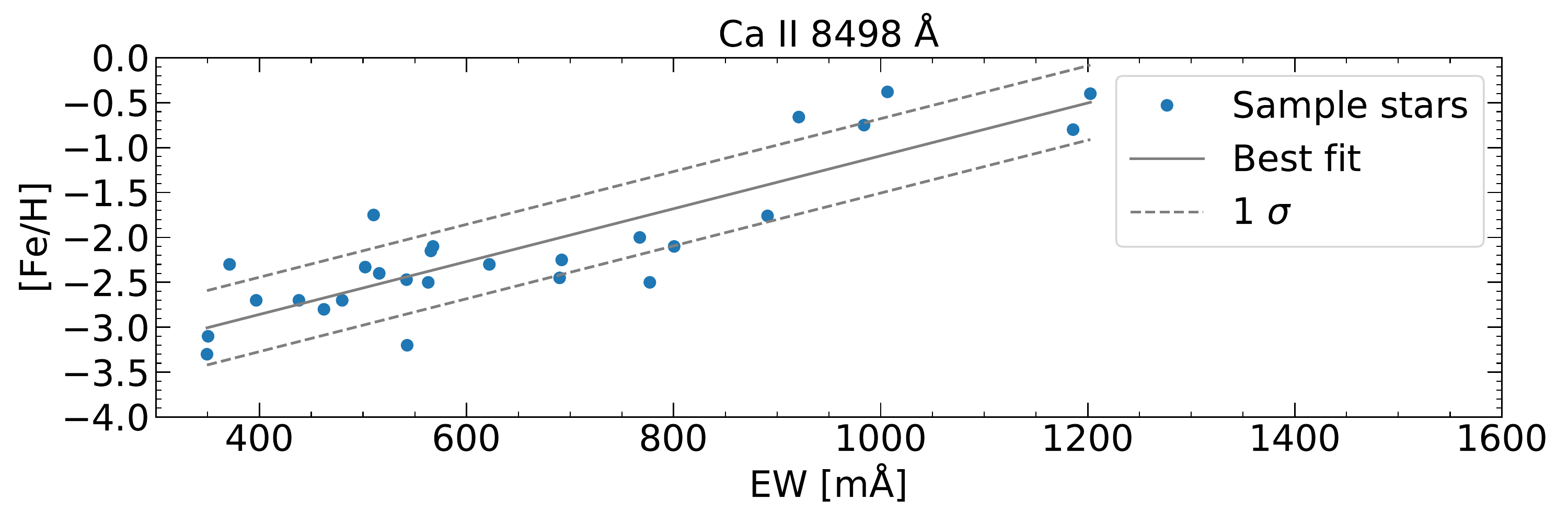}
    \caption{[Fe/H]-EW relation for Ca II for 27 sample (C)EMP stars from Table \ref{A-tab-ew} using the relation from \cite{Wallerstein2012}. A 1$\sigma$ spread around the linear trend is shown with dashed lines.}
         \label{Ca8498}
   \end{figure}
To avoid measuring numerous, blended, at times weak Fe lines that lead to too high [Fe/H], we computed a new empirical formula that fast and efficiently allows for [Fe/H] determinations of large samples of CEMP and C-normal stars. We aim to find a correlation between the metallicity and the EW of another strong spectral feature that can be easily identified, even in metal-poor stars observed at low resolution. We want to use EWs as they do not rely on prior knowledge of the stellar parameters (as abundances do). Hence, the EWs are free from the 1D, LTE assumptions. \citet{Wallerstein2012}  propose a relation between the metallicity and the equivalent width of Ca II at 8498 Å for RR Lyrae stars, see Fig. \ref{Ca8498}. Their relation is given by 

\begin{equation}
\label{rel0}
[\text{Fe/H}]_{RRLyr} = -3.846 (\pm 0.155) + 0.004 (\pm 0.0002) \text{ EW}_{Ca}
.\end{equation}
It has a maximum uncertainty of $\pm$0.25 dex, which is obtained for a typical RR Lyrae star when varying the stated line's intercept and slope by 0.155 and 0.0002, respectively. As the data used in \citet{Wallerstein2012} is not publicly available, quantifying the statistical accuracy of the relation can only be done by varying $\alpha$ and $\beta$ by their standard errors.
However, if this relation is directly applied to our more metal-rich CEMP stars, the uncertainty would go up to $\pm 0.4$\,dex.
The total uncertainty from this relation is larger than what we would derive from EWs of individual (blended) Fe lines, and as seen from Fig.~\ref{Ca8498}, it yields a large scatter when applied to CEMP stars. 

Following \citet{Wallerstein2012}, we plot our sample star metallicities against the EWs of each of the Ca II triplet (CaT) lines at 8498\,Å, 8542\,Å, and 8662\,Å. With a least-squares fit, we find that for Ca II at 8498\,Å, the relation yields a residual standard deviation of 0.313\,dex. We find similar values for the two other Ca II lines' relation and consider the CaT to be unsuitable as a metallicity tracer for CEMP stars. Another reason why the CaT might fail as a tracer is that the strong lines and wings need Voigt or Lorentzian profiles to be fit, therefore deeming Gaussian profiles insufficient. 

Based on this, we sought out to find a similar relation that is suitable for CEMP stars. We first explored a linear relation to see if the most simple relation can provide a fit that yields [Fe/H] estimates, which have a residual standard deviation that is better than 0.25\,dex when starting out with stars with [Fe/H]$<-2.0$. 
The uncertainties of the regression were derived from the sums of squares as follows:

\begin{equation}
\label{ss_xx}
\text{ss}_{xx} = \sum_{i=1}^{n} (x_i - \overline{x})^2
,\end{equation}

\begin{equation}
\label{ss_yy}
\text{ss}_{yy} = \sum_{i=1}^{n} (y_i - \overline{y})^2
, and\end{equation}

\begin{equation}
\label{ss_xy}
\text{ss}_{xy} = \sum_{i=1}^{n} (x_i - \overline{x})(y_i - \overline{y})
.\end{equation}

From the relations in Eq. \ref{ss_xx}, Eq. \ref{ss_yy}, and Eq. \ref{ss_xy}, we can define $s$  as

\begin{equation}
\label{s}
s = \sqrt{\dfrac{\text{ss}_{yy}-\dfrac{\text{ss}_{xy}^2}{\text{ss}_{xx}}}{n-2}}
.\end{equation}

And from Eq. \ref{s}, the standard errors for the regression coefficients $\alpha$ and $\beta$ (line intercept and slope, respectively) are given by

\begin{equation}
\label{SEa}
\text{SE}(\alpha) = s \sqrt{\dfrac{1}{n}+\dfrac{\overline{x}^2}{\text{ss}_{xx}}}
   and\end{equation}

\begin{equation}
\label{SEb}
\text{SE}(\beta) = \dfrac{s}{\sqrt{{\text{ss}_{xx}}}}
.\end{equation}

Using our procedure (see Sect.~\ref{sec:analysis}), we investigate a large number of different elements' absorption lines that we consider to be strong based on cuts in excitation potential (EP $\lesssim3$\,eV) and oscillator strength (log$gf \gtrsim -2.0$). These cuts were made to ensure that the lines were strong and detectable in both metal-poor dwarfs and giants.
The number of stars for which we were able to measure the EWs of elements amounts to Na I (27), Mg I (28), Ca I (9), Ca II (27), Sc II (22), Cr I (44), Ni I (39), and Mn I (12). All of the measured elements can be viewed in Table~\ref{A-tab-ew}. Using spectral synthesis, the number of stars for which we were able to determine elemental abundances are as follows: C (CH) (30), Mg I (30), Sr II (30), Ba II (29), and Eu II (8). We used these abundances to classify the CEMP stars (see Table~\ref{tab-star-para}). 

 \begin{figure}
   \centering
    \includegraphics[width=\hsize]{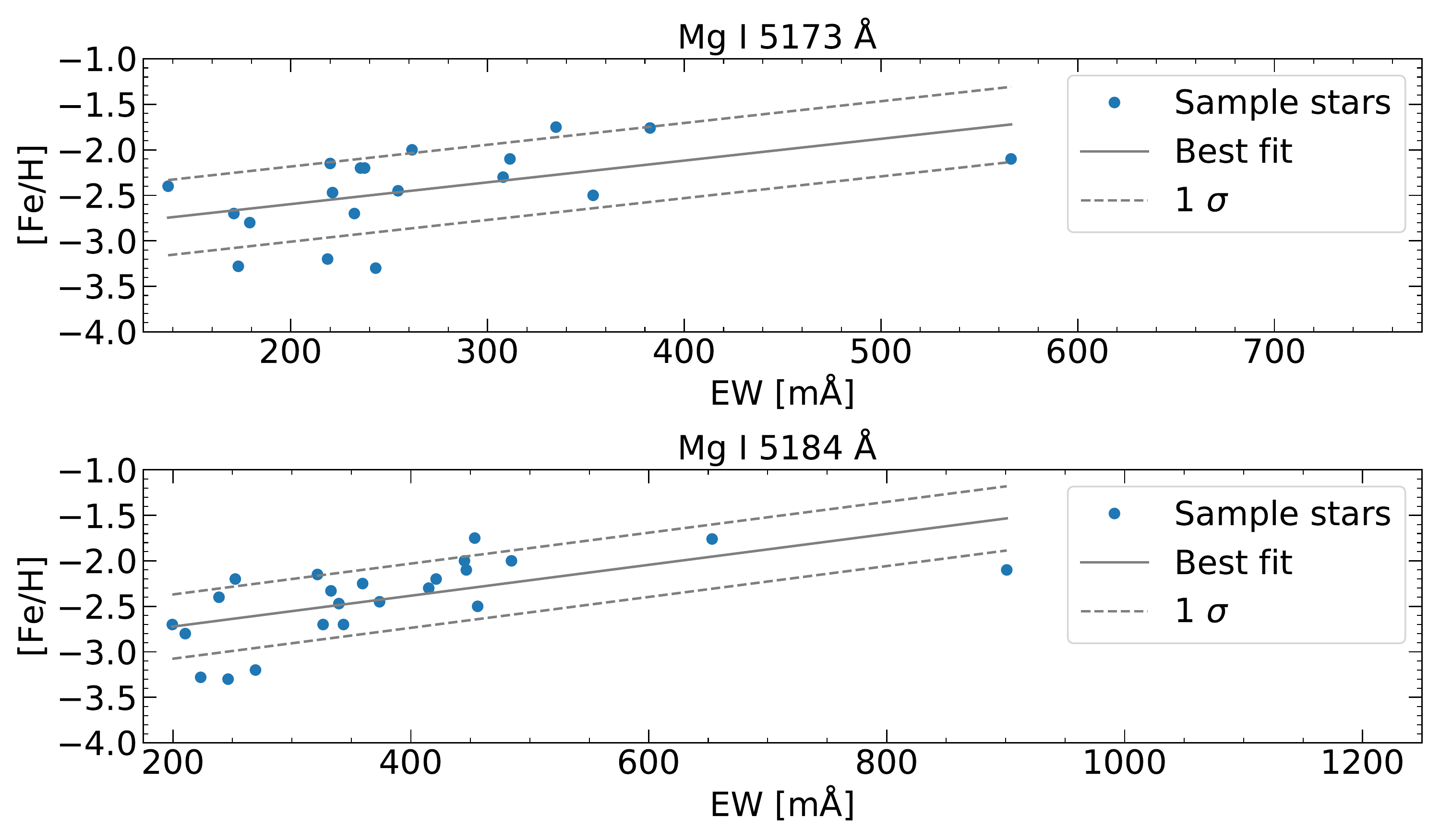}
      \caption{Linear scaling relations of Mg I at 5173\AA\, (19 stars) and 5184\,\AA\ (23 stars) for the  metal-poor stars in Table~\ref{tab-star-para} where the line is not too strong yet ($<1000$ m\AA\,). A scatter larger than 1$\sigma$ (dashed lines) is seen.}
         \label{mg_dobbelt}
   \end{figure}

The Mg I triplet is prominent in the spectra of most stars (at our metallicities and spectral resolution). Due to large molecular bands the Mg I line at 5167\,Å is mostly blended and becomes unusable. Measurements of Mg I  5173\,Å and 5184\,Å provide relations with residual standard deviations of $\pm$ 0.413\,dex and $\pm$ 0.353\,dex, respectively (see Fig. \ref{mg_dobbelt}). They are therefore deemed as inadequate metallicity tracers.

For the Cr I lines at 5844\,Å and 7400\,Å, Mg I at 6173\,Å, Mn I at 4762\,Å, and Sc II at 5239\,Å and 6605\,Å, we discover nearly flat relations with large scatter. Hence, their EWs are degenerate with [Fe/H] (see Fig.~\ref{BadSc}). 
Their residual standard deviations exceed $\pm$ 0.40\,dex, which does not fulfil our targeted accuracy. The Na I lines at 5890\,Å and  5896\,Å, Mg I at 8806 Å, and Ca I at 6162 Å show a linear relation, on the contrary, but their scatter amounts to residual standard deviations exceeding $\pm$ 0.44\,dex, which also excludes them as valid metallicity tracers.

  \begin{figure}
   \centering
    \includegraphics[width=\hsize]{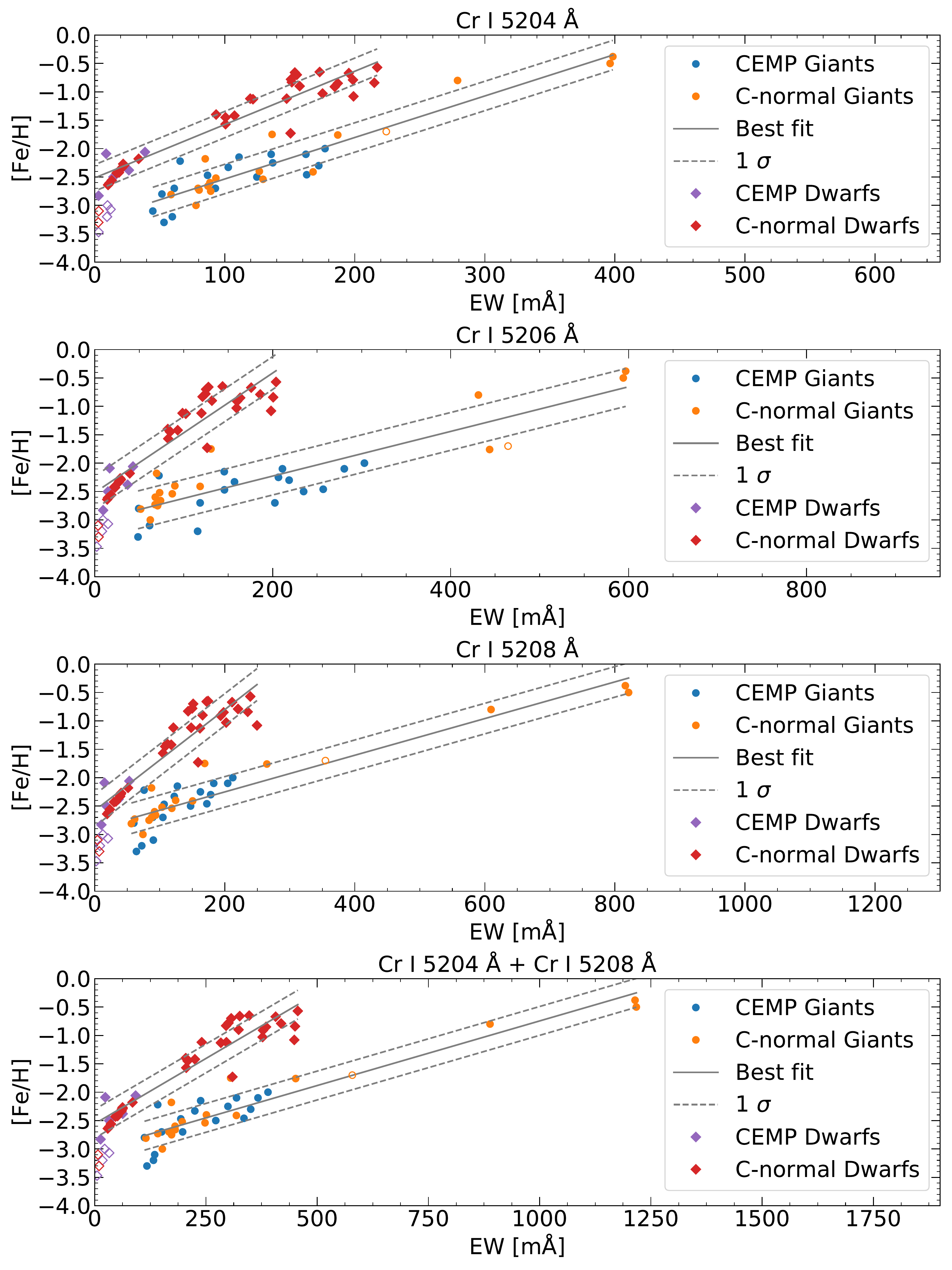}
      \caption{Linear relations between the EWs of Cr I lines and [Fe/H]. The relations shown are for a combined sample of CEMP and C-normal stars from Table \ref{tab:giants} and \ref{tab:dwarfs}. Open symbols have not been included in the linear relations. We note the difference in axes compared to Fig.~\ref{Ca8498}. See legend for details.}
         \label{cr1cr3}
   \end{figure}

However, the relations we derived from the Cr I triplet located at $\sim$ 5204\,Å contain less scatter, except for Cr I at 5206\,Å as this line is blended with other lines (e.g. Y). For these relations, it was necessary to exclude the star \object{HE0516-2515} as an extreme outlier. This is attributed to the poor S/N of 6 at 5000\,Å and in turn the uncertain stellar parameters. We find that by adding the EWs of Cr I at 5204\,Å and Cr I at 5208\,Å, a least-squares fit gives us the the following Cr 5204+5208 relation for giants (see Fig.~\ref{cr1cr3}):

\begin{equation}
\label{rel1}
[\text{Fe/H}]_{Giants} = -3.022 (\pm 0.066) + 0.002 (\pm 0.0002) \text{ EW}_{Cr}
,\end{equation}

where EW is the equivalent width given in mÅ. This Cr I (5204\,Å + 5208\,Å) relation yields a residual standard deviation of 0.254\,dex. This marginally fulfils the requirement on RSD. We note that Eqs.~\ref{rel1}-\ref{rel22} are valid for both CEMP and C-normal stars.

Similarly we find the following for dwarfs:

\begin{equation}
\label{rel12}
[\text{Fe/H}]_{Dwarfs} = -2.565 (\pm 0.071) + 0.005 (\pm 0.0003) \text{ EW}_{Cr}
.\end{equation}

For dwarfs the relation has a residual standard deviation of 0.253\,dex.

\begin{figure}
   \centering
    \includegraphics[width=\hsize]{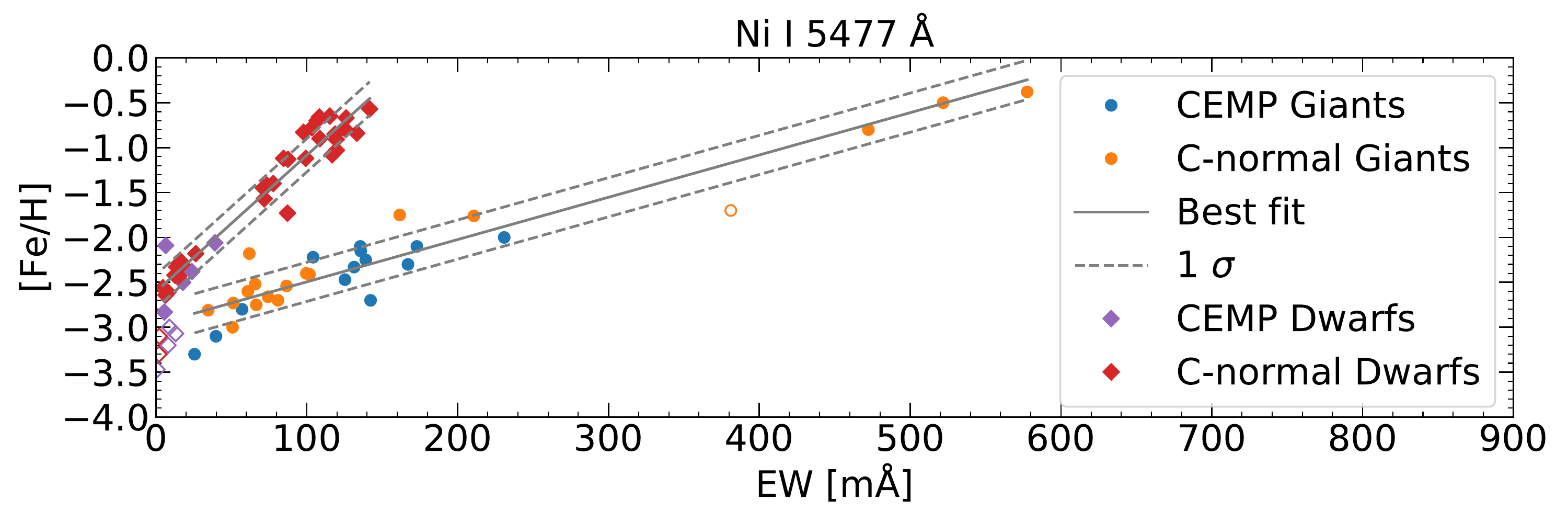}
      \caption{Linear relation between the EW of Ni I line at 5477\,\AA\ and the stellar [Fe/H] for the sample stars from Table \ref{tab:giants} and \ref{tab:dwarfs}.}
         \label{nife}
   \end{figure}

Another spectral feature of interest is Ni I at 5477\,Å. We note that this line is blended with Fe. However, it does not degrade the use of the Ni line (see Sect.~\ref{sec:discussion} for discussions). This element provides our best metallicity tracer. Again, a least-squares fit provides the following relation for giants (Fig.~\ref{nife}):

\begin{equation}
\label{rel21}
[\text{Fe/H}]_{Giants} = -2.965 (\pm 0.058) + 0.005 (\pm 0.0003) \text{ EW}_{Ni}
.\end{equation}

The residual standard deviation for the Ni I relation amounts to 0.218 dex, which is better than conventional methods (see Sect.~\ref{stelpar}). Similarly, for dwarfs we find :

\begin{equation}
\label{rel22}
[\text{Fe/H}]_{Dwarfs} = -2.603  (\pm 0.052) + 0.015 (\pm 0.001) \text{ EW}_{Ni}
.\end{equation}
For dwarfs the residual standard deviation is 0.186\ dex.

The use of the various empirical formulae is illustrated by Table~\ref{NiCrWal} and Table~\ref{tab:correlation}. The use of the Ni tracer is cleaner and closer to the [Fe/H] derived from Fe lines, while the Cr tracers typically agree within $\pm 0.2$\,dex. We also developed tracers for each sub-group of C-normal, CEMP, as well as CEMP and C-normal stars (as listed above) for the dwarfs and giants individually (see Table~\ref{tab:correlation}). If a relation is blindly used for any giant star, for example, with no prior knowledge of C-enhancement, the metallicity is typically accurate to within $\pm 0.15$\,dex.

\begin{table*}
\caption{[Fe/H] obtained with five different methods. Fe I lines, our Ni 5477\,Å relation, our Cr 5204\,Å and Cr 5208\,Å relation, and \citeauthor{Wallerstein2012} Ca 8498\,Å relation, and with ATHOS and SP\_Ace. }
\label{NiCrWal}

\centering

{\begin{tabular}{l  c c c c c c c c c }
\hline \hline \noalign{\smallskip}

Star &  $[$Fe/H$]_\mathrm{FeI lines}$   &       $[$Fe/H$]_\mathrm{NiI 5477Å}$  &       $[$Fe/H$]_\mathrm{CrI 5204Å+5208Å}$ &       $[$Fe/H$]_\mathrm{CaII 8498Å}$ & $[$Fe/H$]_{\rm ATHOS/SP\_\rm{Ace}}$\\
\noalign{\smallskip}
        &               &       Eq. \ref{rel21} &       Eq. \ref{rel1}  &       Eq. \ref{rel0}      \\ \noalign{\smallskip} \hline \noalign{\smallskip}
HE2319-5228     &       -2.8    &       -2.7    &       -2.8    &       -2.0 & -2.6  \\      
HE2250-4229     &       -2.7    &       -2.6    &       -2.7    &       -2.3 &-2.5   \\      
HE2258-4427     &       -2.1    &       -2.1    &       -2.3    &       -0.6& -2.2    \\
HE0448-4806&  -2.3      &   -2.4    &  -2.7       &  -2.4  &  -2.4\\
HE2357-2718& -0.7       &    +0.2      &  -0.3      &   -0.2  & +0.3\\
\hline
\end{tabular}}
\end{table*}

\subsection{Cr and Ni as tracers}
The choice of Cr and Ni was made for several reasons. In the C-free regions, Cr and Ni were among the lines that showed clear absorption features together with elements like Mg and Ca, which unfortunately turned out to be sub-optimal tracers for the C-rich and C-normal dwarfs as well as the giants we study. We wish to stress that we only used the EWs of these two elements, and no abundances are involved in our empirical relations. Thus, no radiative transfer equations or the biasing 1D, LTE assumptions affect this method.

Ni is an excellent tracer, which is also understandable when considering the chemical evolution of Ni \citep[e.g.][]{Kobayashi2019}. As is seen from observations, the [Ni/Fe] is flat with metallicity and has a scatter as low as 0.03 for very metal-poor stars \citep{Reggiani2017}; it only has a slight scatter (0.1-0.2\,dex) around the solar-scaled ratio.

Cr, on the other hand, shows a decreasing [Cr/Fe] below [Fe/H]$=-2.5$ \citep{Bonifacio2009}, reaching down to -0.5 at the lowest metallicities. This sort of trend could propagate into the EWs; however, this does not seem to be the case. In the above-mentioned study, they show that for the less non-LTE biased Cr II abundances, the [Cr II/Fe] trend stays almost flat, and centred around zero with a spread of  $\sim \pm 0.2$\,dex (except from two outliers). This would indicate that the decreasing abundance trend is mainly due to deviations from LTE, which are most strongly present at low metallicities; a similar result is shown in \citet{Reggiani2017}. Hence, the trend in Cr abundance is more so due to the simplified abundances assumptions made when computing Cr abundances, while the EWs seem to follow the metallicity well. In order to settle this, full 3D, non-LTE abundances will need to be computed over a broad metallicity range in order to probe the true nature of the chemical evolution of Cr in the Galaxy.

Hence, selecting two Fe-peak elements to trace the metallicity, seems logical from a nucleosynthetic point of view.\ Additionally, we do not expect to find large deviations from the Fe-scale as a function of (lower) metallicity.

\subsection{Robustness of method}
In order to quantify the robustness of our empirical scaling relations, we derived the stellar parameters and EWs in different ways. For the C-normal stars, the stellar parameters were derived with the automatic code ATHOS \citep{Hanke2018} and these were in good agreement with the temperatures from the IRFM, to within 50-70\,K, and the gravities based on Gaia DR2 values (where the agreement is $\sim \pm 0.05$\,dex). The metallicity obtained measuring EWs is agreeing within $< \pm0.1$\,dex, as shown in Reggiani et al. (submitted). Thus, the impact on the metallicity is low, and the values are consistent.

Following, we re-measured (blindly) the EWs in several stars, which in C-normal stars resulted in $\pm3$\,m\AA\ for the Ni line, and $\sim8$\,m\AA\ for the Cr lines. The uncertainties were larger for the CEMP stars where both cases resulted in $\sim 10-15$\,m\AA . The main reason for which was continuum placement, which is more difficult in CEMP than in C-normal stars. To increase the robustness, automated codes must be able to place the continuum better, which is notoriously hard in cool (CEMP) stars with many molecular bands.

In addition, we used an automated routine \citep[iSpec,][]{2014ascl.soft09006B} to  measure the EWs of Ni and Cr automatically for all dwarf stars considered. The median offset for Ni is 3\,m\AA\ , for Cr(5204\AA) 5\,m\AA\ , and for Cr(5206\AA) 9\,m\AA . Finally, Cr(5208\AA) shows the largest uncertainty of 9\,m\AA,\, which is partly due to blends at higher metallicities and partly due to saturation. These offsets are in overall good agreement with values obtained for CEMP stars.

We computed new empirical relations with the manually and automatically obtained EWs and found the best agreement for Ni. The difference in intercept is $<0.02,$  $<0.01$ in slope, and the scatter is similarly low. For the Cr line at 5204\,\AA,\, the intercept changed slightly more ($<0.03$); however, no change was seen in the slope. Hence, only slight effects on the second decimal were seen for these two lines. The two redder Cr lines were more affected. The intercept changed by 0.1-0.3, while the slopes suffered by only 0.01, which is similar to the blue Cr and the Ni lines. If we vary the EWs by the uncertainties that arise from either continuum placement or from measuring the EWs manually versus automatically, we find that the empirical relation for Ni and the blue Cr line (5204) is robust to within $\sim 0.05$\,dex typically. However, the redder Cr lines are only robust to within $\sim 0.1$\,dex. Hence, for all dwarfs, we recommend using Ni or Cr (5204), and our scaling relations are robust ($\pm0.05$\,dex) and valid in the metallicity range $-3.2$ to 0.

\subsection{Consistency tests}
As a follow-up, we tested the application of the automatic codes SP\_Ace \citep{Boeche2016}\footnote{Version 1.3: http://dc.zah.uni-heidelberg.de/sp\_ace/q/dist/static/} and ATHOS \citep{Hanke2018}\footnote{https://github.com/mihanke/athos} for the giant stars of our sample, shown in Table~\ref{tab-star-para}. While SP\_Ace computes stellar parameters and abundances for stars with [Fe/H]$>-2.4$, ATHOS only yields stellar parameters, but for a broader parameter range. The average error on the metallicity for ATHOS is given by $\sigma _{\mathrm{ATHOS}} = 0.39\, \rm dex$. The given error of SP\_Ace is smaller, which is caused by a different calculation since $\sigma _\mathrm{SP\_Ace} = \frac{1}{N_\mathrm{Fe}}\cdot \sigma=0.1\, \rm dex$. Both codes agree within $\sigma_{\mathrm{[Fe/H]}} = 0.29 \, \rm dex$, $\sigma_{\mathrm{T}} = 277 \, \rm K,$ and $\sigma_{\mathrm{log} g} = 1.16$\,dex for metallicity, temperature, and surface gravity, respectively. The large difference in gravity is driven by five strongly differing giants, where ATHOS operates outside the range of gravities on which it was trained. SP\_Ace did not converge for three stars, even though the metallicity is in the range of validity, as shown in Table~\ref{tab:stellar_pars}. For five stars, SP\_Ace was used to determine the metallicity, but it was unable to determine Cr and Ni abundances. This along with the disagreement within the stellar parameter demonstrates how challenging it is to derive elemental abundances from low-resolution spectra of CEMP stars with automatic codes. 

  \begin{figure*}
   \centering
   \includegraphics[width=\hsize]{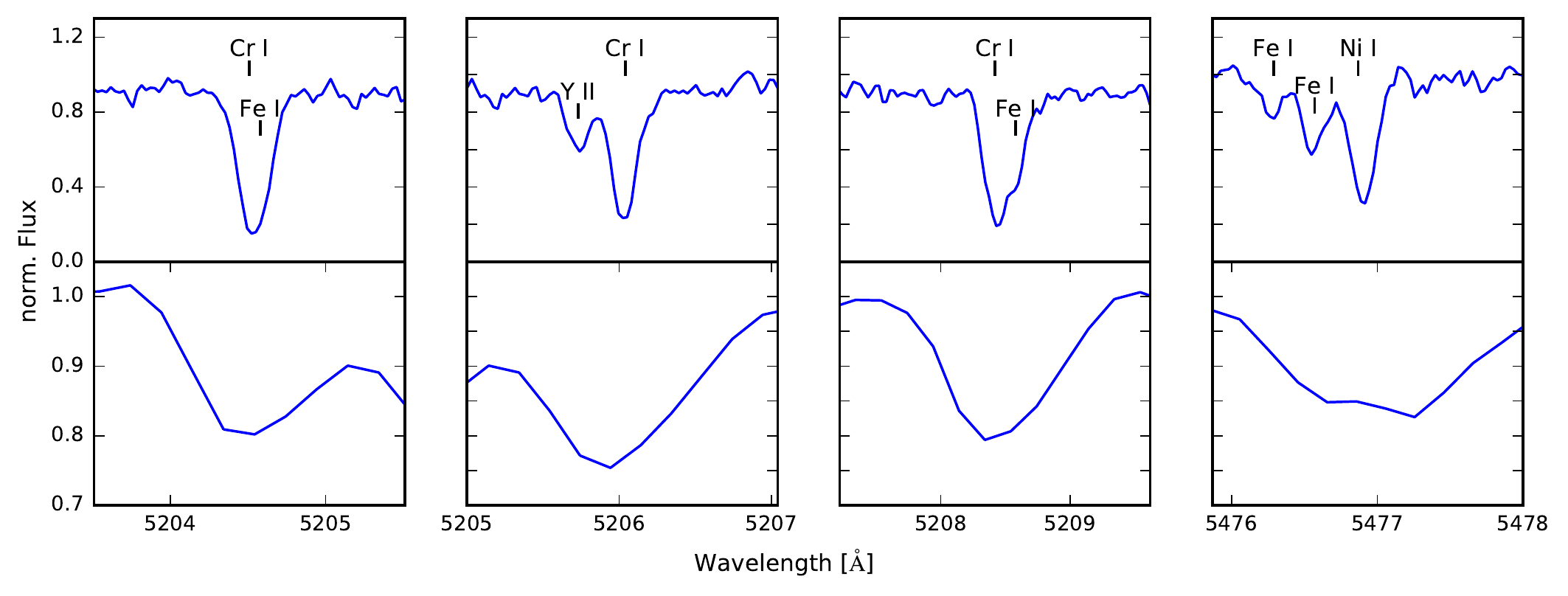}
      \caption{Cr and Ni lines in the high-resolution spectrum of the Sculptor star \object{2MASS J01001255-3343011} with metallicity of [Fe/H$]=\mathrm{-1.62}$ (top) and the low-resolution spectrum of sample star \object{HE 0253-6024} with [Fe/H] =$-2$ (bottom). }
         \label{fig:arcturus}
   \end{figure*}
   
By using Cr and Ni abundances as well as the metallicities determined by SP\_Ace, we calculated the EW of the Cr and Ni lines investigated here. This was done with the EW driver of the spectral synthesis code MOOG \citep[][version of 2014]{Sneden1973} and atomic data, as given in Table~\ref{tab:atomic_data}. Here, MOOG calculates the EW line by line as if it were an isolated feature. 
As seen in Fig.~\ref{fig:arcturus}, the Cr and Ni lines are  heavily blended at the resolution of X-Shooter with either Y or Fe lines.
In many cases, line blends can be well treated, especially if the blends are separated from core to core by more than 3/5*FWHM \citep[full width at half maximum][]{Hansen2015}. However, if the blends are close and of similar strength, they cannot be treated separately; one line's opacity and continuum source function is affected by the other, and the radiative transfer equations need to be solved with this in mind \citep{Blends}. This influence is, however, negligible if the second blending line is much weaker than the first line of interest. 

By inspecting Fig.~\ref{fig:arcturus}, the bluest Cr line poses a problem. In the X-Shooter spectra, the blending separation is $\sim0.5$\,\AA\ and the blending Fe line is closer than this separation and is of similar strength to the Cr line. Hence, treating these two lines completely independently when deriving abundances and EWs may lead to wrong results, and we do not consider the 5204\,\AA\ Cr line any further in this test.

\begin{figure}
   \centering
   \includegraphics[width=\hsize]{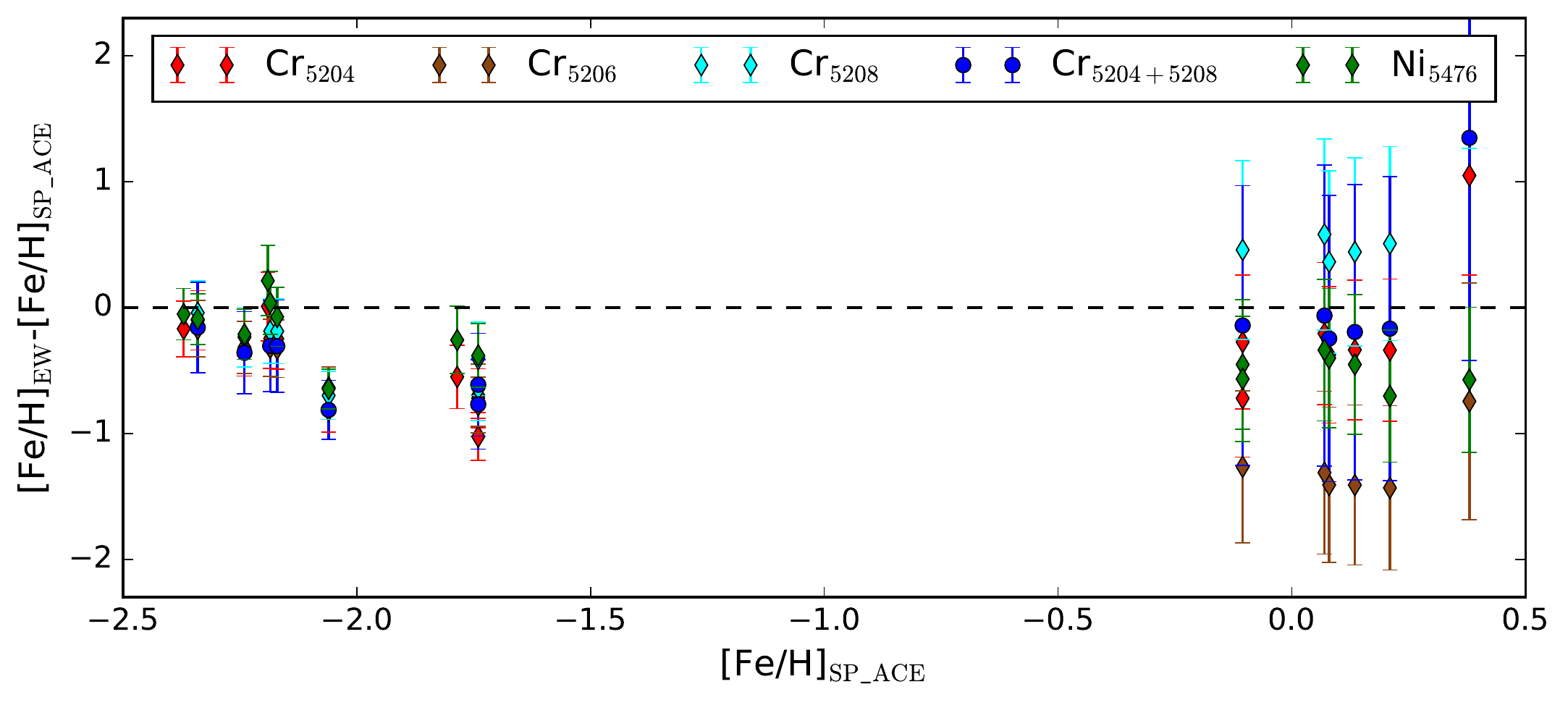}
      \caption{Comparison of metallicities derived with SP\_Ace to metallicities derived from Cr and Ni line EWs for the giants in Table~\ref{tab-star-para}. We note that the EWs in these relations are based on abundances from SP\_Ace. The error bars include 13 stars for Cr and 16 for Ni, see also Table~\ref{tab:stellar_pars}.}
         \label{fig:space_ew_comp}
   \end{figure}

The 5206\,\AA\ Cr line suffers from a Y blend. Even though Y is more than 0.5\,\AA\ from the Cr core, the fact that the slow neutron-capture element, Y, does not scale with [Fe/H] is obvious, and this causes large offsets when applying Cr and Y as an [Fe/H] tracer in the s-process rich CEMP$-s$ stars (see Fig.~\ref{fig:space_ew_comp}). From Fig.~\ref{fig:space_ew_comp}, Ni and Cr 5208 are seen to be the best tracers. We also note that the super-solar [Fe/H] values arise due to the issues automated codes have when dealing with heavily blended, metal-rich stars.
To account for the strong blends and for what we would measure as EW of the blended feature, we added the EW of the blending Fe to the EW of the Cr line. We could only do this for the 5208 Cr and 5477 Ni lines, where the Fe lines are well separated and much weaker than the Cr or Ni lines. We tested the impact of this, and the recovered metallicity is better when we added the Fe EW to the tracer line EW.
The Ni and Cr 5208 (and then 5208+5204)\,\AA\ remain the best [Fe/H] tracers for giants at all metallicities for the sample from Table~\ref{tab-star-para}.
Our scaling relations for giants are valid and robust in the metallicity range $-3.2$ to $-0.5$\,dex for Cr and $-3.2$ to $0$\,dex for Ni.

\section{Discussion}\label{sec:discussion}
To pinpoint where the relations break down, we expanded our sample. We used a large sample of stars, typically C-normal, which are located in dwarf spheroidal galaxies. The spectra were downloaded from the ESO and Keck reduced archive and analysed in a homogeneous manner (Reichert et al. 2019 submitted). Due to the distance to these galaxies, only the brightest, cold giants were observed. These $300$ giant and super-giant stars have similar stellar parameters to the CEMP stars with surface gravities \mbox{$0.37 \, \mathrm{dex} \le \log g \le 2.4 \, \mathrm{dex}$}, temperatures between \mbox{$3700 \, \rm K \le T \le 5300\, \rm K,$} and metallicities between \mbox{$-4.18 \, \mathrm{dex} \le \mathrm{[Fe/H]} \le -0.12\, \mathrm{dex}$}. To derive the stellar parameters, again, we used a combination of SP\_Ace for high metallicities ($\mathrm{[Fe/H]}>-2.3$) and ATHOS for low metallicities, since SP\_Ace is not valid below [Fe/H]=$-2.4$. A detailed description of how stellar parameters are derived is given in  Reichert et al. (2019). The majority of this sample was observed within the HR10/13/14a set-up of FLAMES/GIRAFFE (230 stars), but it also contains UVES (59), X-SHOOTER (1), and HIRES (10) spectra. 

\begin{figure}
   \centering
   \includegraphics[width=\hsize]{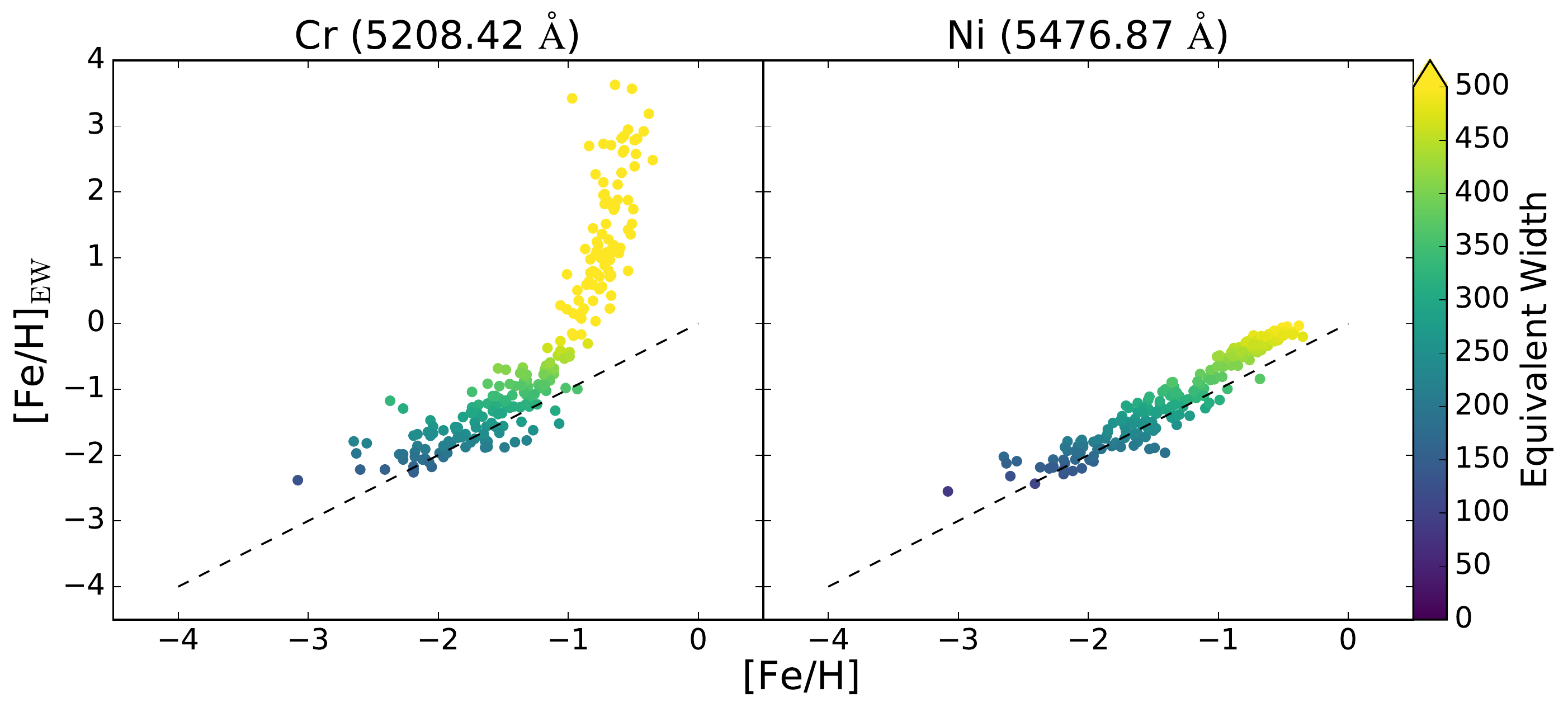}
    \caption{Metallicities from SP\_Ace compared to metallicities derived from EW relations (Cr left, Ni right) in a sample of $230$ C-normal giants and supergiants with GIRAFFE spectra.}
    \label{fig:compdsphGIRAFFE}
\end{figure}
\begin{figure*}
   \centering
   \includegraphics[width=\hsize]{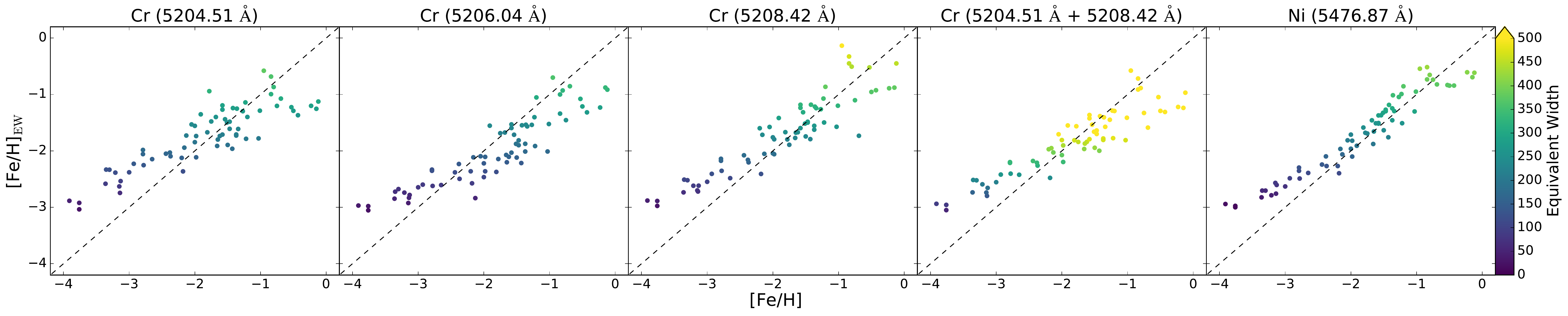}
      \caption{Metallicities from Fe lines (SP\_Ace) compared to metallicities derived from EW relations in a UVES sample of $59$ C-normal giants and supergiants. The line strength is indicated by colour.}
         \label{fig:compdsphUVES}
\end{figure*}

For the FLAMES/GIRAFFE spectra, lines around 5204\,\AA\ are not covered. We, therefore, measured the Cr line at $5409.77\,\AA$ and calculated the EW of the Cr lines that we investigate in this paper. This was done by computing the Cr abundance from the Cr EW at $5409.77\,\AA$ and generalising this to the Cr lines at 5204 -- 5208\,\AA\ as outlined above. We want to stress that calculating the EW in this way, requires knowledge of all stellar parameters. Since Cr is temperature sensitive, this introduces additional uncertainties compared to a direct measurement of the Cr lines at 5200\,\AA . As before, we mimicked the low-resolution spectra by adding EWs of nearby Fe lines, therefore the derived metallicity goes directly into our calculation and may reduce the uncertainty again. Despite the higher resolution of GIRAFFE, the Fe blend of the 5204\,\AA\ Cr line is only 0.07\,\AA\ from the Cr line centre, which is well within FWHM dictated separation that is 0.13\,\AA\ for GIRAFFE. For the redder Cr line, 5206\,\AA , we did not attempt to add up the EW of Y that is close the $\mathrm{Cr}_{5206.04}$ because Y does not trace metallicities and it might introduce biases in s-rich CEMP-s stars. This relation, therefore, turns out to provide metallicities that are too low, as seen in Fig.~\ref{fig:compdsphUVES}. We attribute the large upturn in Fig.~\ref{fig:compdsphGIRAFFE} of the Cr$_\mathrm{5208.51}$ line to saturation of the strong Cr lines (EW$>450$\,m\AA\,) at the highest metallicities tested here. 
This occurs here since we were forced to use Cr abundances derived from other lines in the limited wavelength range, and we use the 5409.77\AA\ Cr line abundance for the lines we study here (5204--5208\AA\ ) to computate their EWs.

Finally, there is a smaller sample of 59 stars that were observed with UVES. Here, the high resolution allowed for a safe deblending down to $\sim 0.035$\,\AA,\, thereby rendering all Cr lines useful tracers. We measured EWs directly and only added the Fe EW to the Ni line in order to mimic the situation in lower-resolution spectra. 
The outcome is shown in Fig.~\ref{fig:compdsphUVES}. The $\mathrm{Cr}_\mathrm{5208.51}$-relation is valid within the boundaries of $50\, \rm m\AA < \mathrm{EW}<450\, \rm m\AA$. This translates to metallicities between $-2.8\, \mathrm{dex} \le \mathrm{[Fe/H]} \le -0.8 \, \mathrm{dex}$. The relation for Ni holds above $\mathrm{EW}>30\, \rm m\AA,$ which translates to $\mathrm{[Fe/H]}>-3.2\, \rm dex$. By applying these boundaries, we derived a residual standard deviation of $\sigma_{5204.51} = 0.53$\,dex, $\sigma_{5206.04} = 0.47$\,dex, $\sigma_{5204.51+5208.42} = 0.44$\,dex, $\sigma_{5208.42} = 0.39$\,dex, and $\sigma_{5476.87} = 0.32$\,dex, as shown in Fig.~\ref{fig:compdsphUVES} and for the FLAMES/GIRAFFE sample (Fig.~\ref{fig:compdsphGIRAFFE}) $\sigma_{5208.42} = 0.29$ and $\sigma_{5476.87} = 0.20$. Again, the 5208 Cr, the combined 5204+5208 Cr, and the Ni line provide the best [Fe/H] traces, but only Ni fulfils our goal with it $\pm0.2$\,dex. There seems to be a slight offset between the trend developed for low-resolution EWs and [Fe/H] and the higher resolution GIRAFFE and UVES trends. We assign the majority of this offset to blending issues related to resolution. It has been shown in earlier studies that [Fe/H] derived from low-resolution studies might differ by $\sim$0.3\,dex compared to a high-resolution spectroscopically derived [Fe/H] value of the same star \citep[see, e.g.][]{Hansen2019}. Based on Fig.~\ref{fig:compdsphGIRAFFE} and \ref{fig:compdsphUVES}, we note that our EW$_{Ni}$ - [Fe/H] relation might be able to circumvent issues regarding
resolution. 

We want to stress that calculating the EW this way, requires knowledge of all stellar parameters and we only appled this method for the GIRAFFE sample (230 stars). Since Cr is temperature sensitive, this introduces additional uncertainties, which for $\mathrm{Cr}_\mathrm{5208.42}$ are 0.23 dex on average (temperature: 0.23 dex, logg: 0.02 dex, [Fe/H]: 0.02 dex), as compared to a direct EW measurement of the Cr lines at 5200\,\AA.  This abundances involvement may in turn also explain part of the offset seen between low and high resolution as mentioned above.

\section{Conclusion}
The CEMP stars are Galactic field stars that are often observed with low-resolution instruments; they are inherently metal poor, contain molecules in their atmosphere, and are valued chemical tracers. When identified, CEMP stars become a challenging and very time-consuming task to chemically analyse. The Fe lines can be minuscule due to low metallicity, or they can be blended with other spectral features as a result of low resolution and strong 
molecular bands.

Here, we set out to improve the spectroscopically derived metallicities by finding an empirical relation that predicts [Fe/H], to a more precise degree than by measuring, the very affected EWs of Fe lines. Hence the goal is to achieve a residual standard deviation from `star to linear prediction' that is better than $0.25\, \rm dex$.

In this study, we tested seven elements that show strong absorption features at low metallicities. Out of these, Cr and in particular Ni turn out to be the best tracers. From a Galactic chemical evolution standpoint, this is not surprising as these two elements show almost flat trends with [Fe/H] \citep{Kobayashi2019, Bonifacio2009} if Cr II lines are used. However, we used the strong Cr I features at $\sim$5200\,\AA\ . From this region, it also becomes evident that spectral blends with Fe lines still pose as valid tracers while, for example, Y formed by the s-process \citep{Bisterzo2014} induces biases in CEMP$-s$ and CEMP$-r/s$ stars.

The empirical relations from Cr ($5208\,\AA$) and Ni ($5477\,\AA$) hold for all CEMP sub-groups as well as C-normal stars, with the caveat that dwarfs and giants need to be separated due to the very different line strength and line behaviour in their atmospheres. However, this is not difficult thanks to the Gaia mission (see Fig.~\ref{GaiaMv}).
The relations are valid in stars with [Fe/H]$\sim -3$  up to [Fe/H]$\sim 0$ with EWs from $\sim5$ to  $\sim 800$\,m\AA,\ which we tested in a sample of $\sim 400$ stars.

Having tested the relations on a broad variety of resolutions (from $R\sim 5000 - 40000$), we see that resolution plays an important role in when and how we can use the EWs of blended lines. Moreover, it is known that when deriving [Fe/H] in a star observed at low and medium resolution versus high resolution, the resulting [Fe/H] may deviate by $>0.3$\,dex \citep[e.g.][]{Hansen2019}. By using the blended Ni-Fe feature at 5477\,\AA\ and by adding their EWs, we may be able to circumvent this resolution driven issue and derive consistent [Fe/H], regardless of resolving power. Finally, by using EWs rather than abundances, we reduce the biases from 1D, LTE assumptions. This is very promising.

Our formulae are valid in the metallicity range $\sim$ $-3<$ [Fe/H] $<0$ and internally robust to within $0.05-0.1$\,dex. Overall, they accurately return metallicities ([Fe/H]) for CEMP stars to within 0.2(5)\,dex for Ni (Cr) and they perform even better for C-normal stars. The robustness and accuracy depend slightly on the trace element and line as well as the stellar evolutionary stage.
These empirical formulae will tremendously ease the [Fe/H] determination in future, deep, low- and medium-resolution surveys that encompass numerous CEMP stars, as these grow in number as metallicity decreases. Having a straightforward way of dealing with and avoiding measuring blended Fe lines in these stars will in turn lower the uncertainty on [Fe/H]. These formulae are a step in that direction.

\begin{acknowledgements}
DS and JSB thank DARK Cosmology Centre and Fonden Dr. N P Wieth-Knudsens Observatorium for support, CJH acknowledges support from the Max Planck Society, and MR support from ERC 677912. HR thanks the PDSE programme (program 88881.132145/2016-01) and JHU. This study was financed in part by the Coordenação de Aperfeiçoamento de Pessoal de Nível Superior - Brasil (CAPES) - Finance Code 001.
We are grateful to Almudena Arcones for useful input and to Anja C. Andersen, Andrew J. Gallagher, and Andreas Koch for scientific discussions and comments on the manuscript.
This work has made use of the SP\_${\rm Ace}$ spectral analysis tool v. 1.3. This work has made use of data from the European Space Agency (ESA) mission {\it Gaia} (\url{https://www.cosmos.esa.int/gaia}), processed by the {\it Gaia} Data Processing and Analysis Consortium (DPAC, \url{https://www.cosmos.esa.int/web/gaia/dpac/consortium}). Funding for the DPAC has been provided by national institutions, in particular the institutions participating in the {\it Gaia} Multilateral Agreement. This research has made use of the Keck Observatory Archive (KOA), which is operated by the W. M. Keck Observatory and the NASA Exoplanet Science Institute (NExScI), under contract with the National Aeronautics and Space Administration. Some of the data presented herein were obtained at the W. M. Keck Observatory, which is operated as a scientific partnership among the California Institute of Technology, the University of California and the National Aeronautics and Space Administration. The Observatory was made possible by the generous financial support of the W. M. Keck Foundation. Furthermore, we made use of the services of the ESO Science Archive Facility under request number mreichert 373937, 376650, 376679, and 377450. This involves data from the program IDs: 092.B-0564, 092.B-0194, 071.B-0641, 098.B-0160, 171.B-0588, 072.D-0245,
089.B-0304, 079.B-0435, 076.B-0391, 080.B-0784, 074.B-0415, 076.B-0146,
180.B-0806, 084.B-0336, 171.B-0520, 065.H-0375, 065.N-0378, 085.D-0536,
072.B-0198, 067.B-0147, 075.D-0075, 082.B-0940, 099.D-0321.
Finally, we thank the anonymous referee for comments.
\end{acknowledgements}

\bibliographystyle{aa}
\bibliography{CEMP}

\newpage

\begin{appendix}
\section{Online material}
\onecolumn

\begin{table*}
 \caption{Sample of giant stars. Stellar ID, measured [Fe/H] in this study, Literature [Fe/H], and EWs of Ni, and three Cr lines. The [Fe/H] references are indicated in parentheses, see table footnote for details.  All EWs have been measured here.}
 \label{tab:giants}
 \centering
\begin{tabular}{l c c | c c c c}
\hline \hline
\noalign{\smallskip}

\noalign{\smallskip}

 & [Fe/H] &[Fe/H]$_\mathrm{Lit}$ &   Ni I     & Cr I     & Cr I     & Cr I         \\
Star         &           & (Reference)  &    5476.9  & 5204.6  & 5206.0  & 5208.4   \\
             & [dex] & [dex] & [m\AA] & [m\AA] &[m\AA] & [m\AA] \\
             \noalign{\smallskip}
\hline
\noalign{\smallskip}
Giants, CEMP\\
\noalign{\smallskip}
\hline
\noalign{\smallskip}

HE0002-1037     &       -2.47   & -2.40 (1)  &          125.3   &       86.8    &       145.8   &       106.8   \\
HE0020-1741     &       -3.30   & -3.60 (1) &   25.6    &       53.3    &       48.7    &       64.2    \\
HE0039-2635     &       -3.20   & -3.20 (1) &    --     &       59.7    &       115.7   &       72.4    \\
HE0059-6540     &       -2.15   & -2.20 (1) &   135.8   &       111.0   &       145.6   &       127.2   \\
HE0253-6024     &       -2.00   & -2.10 (1) &   230.9   &       177.1   &       303.1   &       212.4   \\
HE0317-4705     &       -2.25   & -2.30 (1)  &  139.1   &       136.9   &       206.5   &       162.6   \\
HE0400-2030     &       -2.10   & -2.20 (2) &   135.5   &       135.7   &       211.0   &       183.1   \\
HE0414-0343     &       -2.30   & -2.50 (2) &   167.1   &       172.3   &       218.5   &       178.4   \\
HE0430-4901     &       -3.10   & -3.10 (2) &   39.8    &       44.7    &       61.7    &       90.2    \\
HE0448-4806     &       -2.22   & -2.40 (2) &   104.2   &       65.7    &       72.2    &       76.0    \\
HE0516-2515   &   -2.50   & -2.50$^u$ (2) &     --      & 317.0 & 334.1    & 329.9 \\
HE1431-0245     &       -2.50   & -2.50$^u$ (2) &    --   &     124.7   &       234.8   &       147.5   \\
HE2153-2323     &       -2.46   & -2.40$^u$ (2) &    -- &       163.0   &       256.8   &       172.5   \\
HE2158-5134     &       -2.70   & -3.00 (1) &       --  &       61.2    &       202.5   &       89.3    \\
HE2235-5058     &       -2.70   & -2.70 (2) &   142.3   &       92.8    &       118.5   &       104.9   \\
HE2258-4427     &       -2.10   & -2.10 (1)  &  173.0   &       162.4   &       280.5   &       204.6   \\
HE2319-5228     &       -2.80   & -2.60 (2) &   57.1    &       51.7    &       49.5    &       60.2    \\ 
HE2339-4240     &       -2.33   & -2.30 (1) &   131.4   &       102.7   &       157.1   &       122.4   \\
\hline
\noalign{\smallskip}
Giants, C-normal\\
\noalign{\smallskip}
\hline
\noalign{\smallskip}
BPS CS30312-100  & -- & -2.66 (3)        &   74.3   &  87.1  &  74.0  &  93.7    \\ 
BPS CS31082-001  & --& -2.75 (3)         &   66.5  &  89.2   &  70.7  &  83.6     \\
HD108317        & -- & -2.18 (4)         &       61.9  &  85.0  &  69.6  &  87.4     \\ 
HD122563        & --    & -2.54 (3)      &   86.6  &  129.4 &  87.2     &  118.6   \\ 
HD126587       &        -- & -3.00 (3)   &   50.8  &  77.9  &  62.5  &  74.2    \\ 
HD128279        & --     & -2.52 (5) &   65.8  &  93.2  &  73.0  &  103.5   \\ 
HD165195        & --     &      -2.60 (4) &  60.9  &  88.6  &  68.1  &  92.1    \\ 
HE0315+0000   & --       & -2.73 (6)     &   51.3  &  80.3  &  67.9  &  61.5     \\ 
HE0442-1234   & --       & -2.41 (6)     &   101.8 &  167.9 &  118.5 &  150.6   \\ 
HE1219-0312   & --       & -2.81 (6)     &   34.6  &  58.7   &  51.7  &  56.4     \\ 
HE2250-4229     &       -2.70   &  -- &         80.9    &       79.5    &       71.9    &       87.6    \\
HE2310-4523     &       -2.40   & -- &          99.7    &       126.6   &       90.1    &       124.5   \\
\hline
\noalign{\smallskip}
Giants, [Fe/H]$>-2$\\
\noalign{\smallskip}
\hline
\noalign{\smallskip}
HE0221-3218     &       -0.80   & -0.80 (1)  &          472.3   &       279.0   &       431.2   &       609.5   \\
HE0241-3512     &       -1.76   & -1.80 (2)  &          210.8   &       186.9   &       443.8   &       264.8   \\
HE0408-1733     &       -0.75   & -0.80$^u$(2) &        673.9   &       352.5   &       540.0   &       739.5   \\
HE2138-1616     &       -0.50   & -0.50 (2) &   521.9   &       396.3   &       594.0   &       821.0   \\
HE2141-1441     &       -0.38   & -0.60 (2) &   577.6   &       398.4   &       596.7   &       816.0   \\
HE2144-1832     &       -1.70   & -1.70 (2) &   381.1   &       224.2   &       464.5   &       354.8   \\
HE2357-2718     &       -0.66   & -0.50 (2)&    630.0   &       431.6   &       673.3   &       931.3   \\
HE2358-4640     &       -1.75   & -1.70 (2) &   161.6   &       136.4   &       130.7   &       169.2   \\
\hline
\end{tabular}
\tablefoot{
\tablefoottext{$u$}{Uncertain value.}
}
\tablebib{
(1) \citet{Hansen2019}; (2) \citet{Hansen2016}; (3) \citet{Yong2013}; (4) \citet{Simmerer2004}; (5) \citet{Roederer2014}; (6) \citet{Barklem2005}.}
\end{table*}

\begin{table*}
 \caption{Sample of dwarf stars. Stellar ID, measured [Fe/H], Literature [Fe/H], [C/Fe], and EWs of Ni, and three Cr lines. All EWs and abundances marked by a '*' have been determined here. Spectra of the C-normal dwarfs do not cover the CH (G) band, but are according to literature C-normal.}
 \label{tab:dwarfs}
 \centering
\begin{tabular}{l c c c | c c c c}
\hline \hline
\noalign{\smallskip}
\noalign{\smallskip}
             &   $[$Fe/H$]$      &$[$Fe/H$]_{Lit}$& $[$C/Fe$]$ &   Ni I     & Cr I     & Cr I     & Cr I   \\
Star         &           & (References) &       &  5476.9  & 5204.6  & 5206.0  & 5208.4   \\
         &  [dex]  & [dex] & [dex] & [m\AA] & [m\AA] &[m\AA] & [m\AA] \\
\noalign{\smallskip}
\hline
\noalign{\smallskip}
Dwarfs, CEMP \\ 
\noalign{\smallskip}
\hline
\noalign{\smallskip}
BPS CS22881-036  &      --      & -2.06 (7) &  1.96      &  39.1   &  38.7  &  43.1  &  53.3   \\ 
HE0338-3945    &        --      & -2.38 (8) & 2.13       &   23.8  &  26.3  &  37.0  &  36.7   \\ 
HE0450-4902   & --      & -3.07 (9) & 2.03       &    13.1  &  12.3  &  15.0  &  20.3   \\ 
SDSSJ0912+0216 &        --      & -2.50 (10)     & 2.17  &  17.8  &  14.4  &  14.9  &  18.0          \\ 
SDSSJ1036+1212 &        --      & -3.20 (10)     & 1.47  &  8.1   &  9.5   &  8.2        &  8.1     \\ 
SDSSJ1349-0229 &        --      & -3.00 (10)     & 2.82  &  9.0   &  9.9   &  9.3        &  12.6   \\ 
BPS CS22949-0008        &       --      &       -2.09 (12), (13)        &       1.51         &       6.5     &       8.8     &       16.8    &       15.1    \\
HE1148-0037b    &       --      &       -3.47 (12), (14)        &       0.80         &       0.8     &       3.0     &       2.3     &       2.4     \\
BPS CS29514-0007        &       --      &       -2.83 (12), (15)        &       0.89         &       5.6     &       3.0     &       9.5     &       10.3    \\
\noalign{\smallskip}
\hline                          
\noalign{\smallskip}
Dwarfs, C-normal        \\                              
\noalign{\smallskip}
\hline                          
\noalign{\smallskip}
BD+203603     & --       & -2.18 (11)&  0.08* &   26.4    &  33.8   &   39.4    &  51.3    \\ 
BD+241676     & --       & -2.43 (11)&  0.35* &   15.1    &  16.9   &   22.9    &  29.8    \\ 
BD+262621     & --       & -2.61 (11)&  0.50* &   8.0     &  11.3   &   15.0    &  21.3    \\ 
BD-043208     & --       & -2.33 (11)& 0.28*     &   13.9    &  21.3   &   27.3    &  39.1    \\ 
BD-133442     & --       & -2.64 (11)& 0.50*     &   6.5     &  10.3   &   14.1    &  19.1    \\ 
BPS CS22943-0095  &     --       & -2.30 (11)& 0.40*     &   17.4    &  22.5   &   27.8    &  38.5 \\ 
CD-7101234    & --       & -2.42 (11)& 0.24*     &   12.6    &  18.7   &   22.1    &  32.7    \\ 
G126-052      & --       & -2.27 (11)& 0.27*     &   16.7    &  21.9   &   28.7    &  40.7    \\ 
HD338529      & --       & -2.29 (11)& 0.38*     &   15.1    &  22.7   &   29.4    &  40.1    \\ 
HD340279      & --       & -2.56 (11)& 0.44*     &   4.6     &  13.1   &   18.6    &  23.3    \\ 
BD-043208       &       --      &       -2.33 (12)      &       --      &       13.9    &       21.3    &       27.3    &       39.1    \\
HD111980        &       --      &       -1.13 (12)      &       --      &       87.5    &       121.7   &       102.4   &       161.9   \\
BD-01306        &       --      &       -0.83 (12)      &       --      &       97.9    &       151.6   &       121.2   &       143.7   \\
G66-51  &       --      &       -0.79 (12)      &       --      &       125.1   &       198.6   &       186.1   &       220.4   \\
HD204155        &       --      &       -0.66 (12)      &       --      &       108.3   &       153.9   &       127.8   &       172.1   \\
HD134088        &       --      &       -0.90 (12)      &       --      &       108.7   &       157.6   &       131.7   &       165.9   \\
HD224930        &       --      &       -0.91 (12)      &       --      &       119.6   &       184.6   &       160     &       193.7   \\
HD17820 &       --      &       -0.70 (12)      &       --      &       106.8   &       155.4   &       125.2   &       151.5   \\
CD-253416       &       --      &       -0.57 (12)      &       --      &       141.7   &       217.2   &       203.8   &       239.3   \\
HD120559        &       --      &       -1.03 (12)      &       --      &       119.7   &       175.2   &       159.5   &       202.2   \\
G188-30 &       --      &       -1.73 (12)      &       --      &       87.2    &       150.6   &       126.4   &       159.1   \\
HD149414        &       --      &       -1.08 (12)      &       --      &       116.8   &       199     &       198.2   &       249.6   \\
G161-73 &       --      &       -1.42 (12)      &       --      &       73.7    &       107.5   &       93.3    &       117.7   \\
BD-094604       &       --      &       -1.57 (12)      &       --      &       71.8    &       100.6   &       82.6    &       105.1   \\
BD+114725       &       --      &       -0.85 (12)      &       --      &       118.7   &       186.9   &       163.5   &       198.2   \\
HD345957        &       --      &       -1.45 (12)      &       --      &       71.2    &       100.6   &       84.7    &       108.4   \\
HD88725 &       --      &       -0.65 (12)      &       --      &       115.3   &       173     &       143.7   &       174.3   \\
HD106038        &       --      &       -1.40 (12)      &       --      &       77.8    &       93.4    &       82.2    &       111.3   \\
HD134113        &       --      &       -0.78 (12)      &       --      &       103.4   &       150.9   &       124.5   &       150     \\
HD64606 &       --      &       -0.84 (12)      &       --      &       133.2   &       215     &       200.6   &       235.3   \\
BD-086177       &       --      &       -0.67 (12)      &       --      &       126.1   &       195.4   &       175.9   &       211.3   \\
BD+29366        &       --      &       -1.12 (12)      &       --      &       99.3    &       147.6   &       120     &       148.4   \\
HD179626        &       --      &       -1.12 (12)      &       --      &       84.6    &       119.6   &       98.7    &       121     \\

\noalign{\smallskip}
G64-12  &       -3.30   & -3.30 &       0.5*    &       2.0     &       2.8     &       4.4     &       7.0     \\
G64-37  &       -3.10   & -3.05 &       $<0.7*$&        2.2     &       3.1     &       3.8     &       4.2     \\

\hline                                          
\end{tabular}
\tablefoot{
\tablefoottext{*}{Values derived in this study.}

}
\tablebib{
(7) \citet{Preston2001}; (8) \citet{Jonsell2006};  (9) \citet{THansen2015}; (10) \citet{Masseron2010}; (11) \citet{Reggiani2017}; (12) Reggiani et al. in prep; (13) \citet{Masseron_2012}; (14) \citet{Barklem:2005vf}; (15) \citet{Roederer_2014}.}

\end{table*}

\begin{sidewaystable*}
\caption{Equivalent width of strong lines. This includes lines we could measure in most stars, while lines only measured in some stars were of low or no interest. }
\label{A-tab-ew}
\centering
\resizebox{\textheight}{!}{\begin{tabular}{l  c c c c c c c c c c c c c c c c c c c c c c c c c c c c c c c c c c c c c c c c c c c c c c c}
\hline
\noalign{\smallskip}
\hline
\noalign{\smallskip}
\noalign{\smallskip}
                &       Na I    &       Na I            &       Mg I    &       Mg I       &       Mg I    &       Mg I    &       Ca I    &       Ca I    &       Ca I       &       Ca II & Ca II & Ca II   &       Sc II   &       Sc II   &       Sc II      &       Sc II   &       Sc II   &       Cr I    &       Cr I    &       Cr I       &       Cr I    &       Cr I    &       Ni I    &       Ni I    &       Mn I       \\ 
                &       5889.95 &       5895.92         &       5172.69 &       5183.61 &       5528.32 &       8806.70 &       6102.72 &       6122.22 &       6162.17 &       8498.02 & 8542.09 &       8662.14 &       5239.81 &       5526.78 &       5641.00 &       5667.16 &       6604.58 &       5204.57 &       5206.02 &       5208.41 &       5844.6  &       7400.18 &       3492.96 &       5476.87 &       4762.32 \\
                \noalign{\smallskip}
                \hline
                \noalign{\smallskip}
                \noalign{\smallskip}
                EP [eV] & 0.0 & 0.0 & 2.71 & 2.72 & 4.35 & 4.35 & 1.88 & 1.89 & 1.90 & 1.69 & 1.70 & 1.69 & 1.46 & 1.77 & 1.50 & 1.50 & 1.36 & 0.94 & 0.94 & 0.94 & 3.01 & 2.90 & 0.11 & 1.83 & 1.94 \\
                log $gf$ [dex] & 0.11 & -0.19 & -0.39 & -0.17 & -0.50 & -0.34 & -0.79 & -0.32 & -0.09 & -1.32 & -0.36 & -0.62 & -0.77 & 0.02 & -1.13 & -1.31 & -1.31 & -0.21 & 0.02 & 0.16 & -1.76 & -0.11 & -0.27 & -0.89 & -2.43 \\
                \noalign{\smallskip}
                \hline
                \noalign{\smallskip}
Star            &               &                       &               &               &               &               &               &               &               &               &               &               &               &               &               &               &               &               &               &               &               &               &               &               \\ 
HE0002--1037    &       514.1   &       414.7   &       221.2   &       339.5   &       167.8   &       143.7   &       --      &       106.6   &       --      &       542     &       --      &       835.3   &       68.4    &       106.5   &       30.9    &       --      &       29.6    &       86.8    &       145.8   &       106.8   &       30.3    &       57.5    &       120.3   &       125.3   &       92.8    \\
HE0020--1741    &       181.5   &       165.9   &       243.2   &       246.3   &       --      &       122.2   &       --      &       --      &       34.7    &       349.3   &       --      &       510.7   &       --      &       --      &       --      &       --      &       --      &       53.3    &       48.7    &       64.2    &       --      &       --      &       138.6   &       25.6    &       --      \\
HE0039--2635    &       232.6   &       192     &       218.7   &       269.4   &       176.3   &       166.6   &       --      &       --      &       107.3   &       542.7   &       --      &       796.9   &       64.4    &       --      &       35      &       29.5    &       35.3    &       59.7    &       115.7   &       72.4    &       --      &       --      &       97.1    &       --      &       84.4    \\
HE0059--6540    &       264.3   &       214.8   &       220.1   &       321.5   &       102.3   &       144     &       66.4    &       92.3    &       100.9   &       565.5   &       --      &       845.6   &       50.4    &       74.1    &       --      &       --      &       22.5    &       111     &       145.6   &       127.2   &       23      &       19.7    &       138.6   &       135.8   &       77.8    \\
HE0221--3218    &       1206.5  &       827.3   &       1791.9  &       1847.6  &       371.5   &       507.4   &       452.5   &       307.9   &       308     &       1185.9  &       --      &       2042.1  &       --      &       --      &       --      &       --      &       --      &       279     &       431.2   &       609.5   &       --      &       --      &       --      &       472.3   &       --      \\
HE0241--3512    &       357.4   &       308.5   &       382.6   &       653.2   &       --      &       271.3   &       --      &       --      &       --      &       890.7   &       --      &       1195    &       --      &       --      &       --      &       --      &       68.1    &       186.9   &       443.8   &       264.8   &       --      &       --      &       448.8   &       210.8   &       --      \\
HE0253--6024    &       383.6   &       296.8   &       261.6   &       445.1   &       --      &       236.5   &       137.8   &       147.6   &       187     &       767.3   &       --      &       1176.8  &       92.3    &       --      &       69.3    &       76.6    &       --      &       177.1   &       303.1   &       212.4   &       62.1    &       --      &       --      &       230.9   &       108.2   \\
HE0317--4705    &       253.7   &       217.2   &       --      &       359.4   &       187.4   &       184.2   &       98.5    &       142.6   &       151.2   &       691.9   &       --      &       969.4   &       65.4    &       166.9   &       --      &       31.8    &       --      &       136.9   &       206.5   &       162.6   &       --      &       59.4    &       --      &       139.1   &       112.6   \\
HE0400--2030    &       300.6   &       237.3   &       311.4   &       446.6   &       --      &       169.7   &       --      &       --      &       --      &       567.6   &       --      &       730.2   &       25.5    &       --      &       --      &       --      &       30.5    &       135.7   &       211     &       183.1   &       --      &       --      &       --      &       135.5   &       47.7    \\
HE0408--1733    &       1162.7  &       837.7   &       1776.2  &       1783.9  &       --      &       575.2   &       --      &       --      &       --      &       983.9   &       2873.2  &       2464.1  &       --      &       --      &       --      &       --      &       --      &       352.5   &       540     &       739.5   &       --      &       --      &       --      &       673.9   &       --      \\
HE0414--0343    &       422.8   &       324.6   &       307.9   &       415     &       --      &       168.3   &       --      &       --      &       --      &       622     &       --      &       1153.5  &       72.3    &       --      &       --      &       --      &       56.1    &       172.3   &       218.5   &       178.4   &       --      &       --      &       --      &       187     &       56.1    \\
HE0430--4901    &       98.6    &       111.3   &       173.3   &       223.3   &       --      &       103.3   &       --      &       --      &       --      &       350.4   &       --      &       536.2   &       15.8    &       --      &       --      &       --      &       --      &       44.7    &       61.7    &       90.2    &       --      &       --      &       --      &       39.8    &       --      \\
HE0448--4806    &       200.8   &       174     &       235.5   &       252.4   &       --      &       124.4   &       --      &       --      &       --      &       371.1   &       --      &       621.9   &       31.1    &       --      &       --      &       --      &       --      &       65.7    &       72.2    &       76      &       --      &       --      &       --      &       104.2   &       --      \\
HE0516--2515    &       374.9   &       289     &       353.7   &       456.2   &       --      &       330.1   &       --      &       --      &       --      &       777     &       --      &       1376.6  &       --      &       --      &       --      &       --      &       --      &       317     &       334.1   &       329.9   &       --      &       --      &       --      &       --      &       --      \\
HE1431--0245    &       207.9   &       194.8   &       237.5   &       421.3   &       --      &       205.1   &       --      &       --      &       --      &       563     &       --      &       1053.6  &       82.9    &       --      &       --      &       --      &       --      &       124.7   &       234.8   &       147.5   &       --      &       --      &       --      &       --      &       --      \\
HE2138--1616    &       1723.7  &       1191.2  &       2058.7  &       2237.4  &       --      &       527.6   &       --      &       --      &       --      &       1202.5  &       --      &       --      &       25.8    &       --      &       --      &       --      &       --      &       396.3   &       594     &       821     &       --      &       --      &       --      &       521.9   &       --      \\
HE2141--1441    &       1198.5  &       960.6   &       1923.2  &       1895.3  &       --      &       563     &       --      &       --      &       --      &       1006.6  &       1281.9  &       2461.1  &       115.9   &       --      &       --      &       --      &       --      &       398.4   &       596.7   &       816     &       --      &       --      &       --      &       577.6   &       --      \\
HE2144--1832    &       --      &       --      &       566.1   &       900.9   &       --      &       --      &       --      &       --      &       --      &       --      &       --      &       --      &       78.2    &       --      &       --      &       --      &       --      &       224.2   &       464.5   &       354.8   &       --      &       --      &       --      &       381.1   &       --      \\
HE2153--2323    &       308.1   &       270.3   &       254.5   &       373.7   &       --      &       221.7   &       --      &       --      &       --      &       689.9   &       --      &       1156.1  &       119.5   &       --      &       --      &       --      &       --      &       163     &       256.8   &       172.5   &       --      &       --      &       --      &       --      &       --      \\
HE2158--5134    &       474.6   &       390     &       --      &       326.2   &       363.4   &       170.7   &       --      &       226.8   &       189.8   &       438.1   &       --      &       667.3   &       71.8    &       --      &       --      &       --      &       --      &       61.2    &       202.5   &       89.3    &       54.7    &       32.2    &       --      &       --      &       117.6   \\
HE2235--5058    &       206.7   &       182     &       232.3   &       343.4   &       --      &       131.5   &       --      &       --      &       --      &       479.9   &       --      &       843.1   &       45.6    &       --      &       --      &       --      &       --      &       92.8    &       118.5   &       104.9   &       --      &       --      &       --      &       152.5   &       --      \\
HE2250--4229    &       178.1   &       141.4   &       171.1   &       199.5   &       --      &       119.7   &       --      &       --      &       --      &       396.8   &       --      &       671.4   &       9.7     &       --      &       --      &       --      &       --      &       79.5    &       71.9    &       87.6    &       --      &       --      &       --      &       80.9    &       --      \\
HE2258--4427    &       266.8   &       209     &       --      &       484.6   &       --      &       205.7   &       --      &       226.8   &       189.8   &       800.5   &       --      &       1183.3  &       101.8   &       --      &       --      &       90.4    &       67.2    &       162.4   &       280.5   &       204.6   &       58.4    &       87.9    &       --      &       173     &       135.3   \\
HE2310--4523    &       184.4   &       158.8   &       137.7   &       238.7   &       --      &       141.7   &       --      &       --      &       --      &       515.7   &       --      &       892     &       14.5    &       --      &       --      &       --      &       --      &       126.6   &       90.1    &       124.5   &       --      &       --      &       --      &       99.7    &       26      \\
HE2319--5228    &       274.7   &       232.3   &       179.2   &       210.4   &       --      &       90.4    &       --      &       --      &       --      &       462.3   &       --      &       754     &       18.8    &       --      &       --      &       --      &       --      &       51.7    &       49.5    &       60.2    &       --      &       --      &       --      &       57.1    &       --      \\
HE2339--4240    &       305.1   &       255.9   &       --      &       332.9   &       134     &       139.1   &       --      &       112.9   &       96.5    &       502.1   &       --      &       820     &       51.1    &       102.8   &       --      &       --      &       22.3    &       102.7   &       157.1   &       122.4   &       24.3    &       17.2    &       --      &       131.4   &       81.7    \\
HE2357--2718    &       2832.2  &       1900.6  &       2441.2  &       2581.2  &       --      &       669.1   &       --      &       --      &       --      &       920.9   &       3213.6  &       2349.9  &       --      &       --      &       --      &       --      &       --      &       431.6   &       673.3   &       931.3   &       --      &       --      &       --      &       630     &       --      \\
HE2358--4640    &       234.1   &       195.6   &       334.8   &       453.7   &       --      &       197.7   &       --      &       --      &       --      &       510.3   &       --      &       1182.7  &       27.5    &       --      &       --      &       --      &       --      &       136.4   &       130.7   &       169.2   &       --      &       --      &       --      &       161.6   &       60.4    \\
 \hline

\end{tabular}}

\end{sidewaystable*}

\begin{table}
\caption{Atomic data for the Cr, Ni, and Fe lines.}           
\label{tab:atomic_data}     
\centering                   
\begin{tabular}{cccc}
\hline\hline       
\noalign{\smallskip}

Wavelength      &       Species &       EP      &        log\textit{gf}\\ 

 [\AA]  &  &    [eV]    &       [dex]   \\ 
\noalign{\smallskip}
\hline                      
\noalign{\smallskip}
5204.51 & 24.0    & 0.941 & -0.19 \\
5206.04 & 24.0    & 0.941 &  0.02 \\
5208.42 & 24.0    & 0.941 &  0.17 \\   
5409.77 & 24.0    & 1.029 & -0.67 \\   
5204.58 & 26.0    & 0.087 & -4.33 \\
5208.59 & 26.0    & 3.239 & -0.98 \\
5476.29 & 26.0    & 4.140 & -0.94 \\
5476.57 & 26.0    & 4.100 & -0.45 \\
5476.87 & 28.0    & 1.825 & -0.89 \\

\hline             
\end{tabular}
\end{table}

\begin{table}
\caption{Correlations for various sub-groups of stars. Sub-group, evolutionary stage, element and line, slope ($\beta$), slope standard error (SE\,($\beta$)), line intercept ($\alpha$), line intercept standard error (SE\,($\alpha$)), and residual standard deviation (RSD).}        
\label{tab:correlation}      
\centering                     
\begin{tabular}{l c c c c c c c}
\hline\hline       
\noalign{\smallskip}
Sub-group & Stellar evolution & Element/line & Slope ($\beta)$\, X & SE\,($\beta$) & + Intercept ($\alpha$) & SE\,($\alpha$) & RSD \\
\noalign{\smallskip}
\hline
\noalign{\smallskip}

CEMP + C-normal & Dwarf & Cr 5204 & 0.009 & $\pm 0.001$ & -2.526 & $\pm 0.063$ & 0.232 \\
CEMP + C-normal & Dwarf & Cr 5206 & 0.011 & $\pm 0.001$ & -2.557 & $\pm 0.082$ & 0.289 \\
CEMP + C-normal & Dwarf & Cr 5208 & 0.009 & $\pm 0.001$ & -2.592 & $\pm 0.081$ & 0.281 \\
CEMP + C-normal & Dwarf & Cr 5204 + 5208 & 0.005 & $\pm 0.0003$ & -2.565 & $\pm 0.071$ & 0.253 \\
CEMP + C-normal & Dwarf & Ni 5477 & 0.015 & $\pm 0.001$ & -2.603 & $\pm 0.052$ & 0.186 \\
\hline
\noalign{\smallskip}
CEMP + C-normal & Giant & Cr 5204 & 0.007 & $\pm 0.001$ & -3.267 & $\pm 0.082$ & 0.259 \\
CEMP + C-normal & Giant & Cr 5206 & 0.004 & $\pm 0.0004$ & -3.014 & $\pm 0.089$ & 0.333 \\
CEMP + C-normal & Giant & Cr 5208 & 0.003 & $\pm 0.0002$ & -2.898 & $\pm 0.063$ & 0.296 \\
CEMP + C-normal & Giant & Cr 5204 + 5208 & 0.002 & $\pm 0.0002$ & -3.022 & $\pm 0.066$ & 0.254 \\
CEMP + C-normal & Giant & Ni 5477 & 0.005 & $\pm 0.0003$ & -2.965 & $\pm 0.058$ & 0.218 \\
\hline

\noalign{\smallskip}
C-normal & Dwarf & Cr 5204 & 0.009 & $\pm 0.001$ & -2.516 & $\pm 0.074$ & 0.232 \\
C-normal & Dwarf & Cr 5206 & 0.011 & $\pm 0.011$ & -2.520 & $\pm 0.098$ & 0.296 \\
C-normal & Dwarf & Cr 5208 & 0.009 & $\pm 0.001$ & -2.581 & $\pm 0.099$ & 0.289 \\
C-normal & Dwarf & Cr 5204 + 5208 & 0.005 & $\pm 0.0003$ & -2.556 & $\pm 0.086$ & 0.257 \\
C-normal & Dwarf & Ni 5477 & 0.015 & $\pm 0.001$ & -2.583 & $\pm 0.057$ & 0.173 \\
\hline
\noalign{\smallskip}
C-normal & Giant & Cr5204 & 0.007 & $\pm 0.001$ & -3.258 & $\pm 0.111$ & 0.259 \\
C-normal & Giant & Cr5206 & 0.004 & $\pm 0.0004$ & -2.871 & $\pm 0.101$ & 0.299 \\
C-normal & Giant & Cr5208 & 0.003 & $\pm 0.0002$ & -2.838 & $\pm 0.078$ & 0.239 \\
C-normal & Giant & Cr5204 + 5208 & 0.002 & $\pm 0.0002$ & -2.967 & $\pm 0.083$ & 0.236 \\
C-normal & Giant & Ni 5477 & 0.005 & $\pm 0.0003$ & -2.887 & $\pm 0.067$ & 0.200 \\
\hline
\noalign{\smallskip}
CEMP & Dwarf & Cr 5204 & 0.014 & $\pm 0.014$ & -2.620 & $\pm 0.226$ & 0.290 \\
CEMP & Dwarf & Cr 5206 & 0.013 & $\pm 0.01$ & -2.693 & $\pm 0.269$ & 0.288 \\
CEMP & Dwarf & Cr 5208 & 0.011 & $\pm 0.008$ & -2.653 & $\pm 0.255$ & 0.294 \\
CEMP & Dwarf & Cr 5204 + 5208 & 0.006 & $\pm 0.005$ & -2.640 & $\pm 0.242$ & 0.292 \\
CEMP & Dwarf & Ni 5477 & 0.012 & $\pm 0.011$ & -2.591 & $\pm 0.254$ & 0.315 \\
\hline
\noalign{\smallskip}
CEMP & Giant & Cr 5204 & 0.007 & $\pm 0.001$ & -3.204 & $\pm 0.17$ & 0.272 \\
CEMP & Giant & Cr 5206 & 0.003 & $\pm 0.001$ & -3.063 & $\pm 0.173$ & 0.306 \\
CEMP & Giant & Cr 5208 & 0.006 & $\pm 0.001$ & -3.282 & $\pm 0.183$ & 0.270 \\
CEMP & Giant & Cr 5204 + 5208 & 0.003 & $\pm 0.001$ & -3.258 & $\pm 0.176$ & 0.268 \\
CEMP & Giant & Ni 5477 & 0.006 & $\pm 0.001$ & -3.211 & $\pm 0.153$ & 0.222 \\

\hline
\noalign{\smallskip}

\end{tabular}
\end{table}

\begin{figure}
   \centering
    \includegraphics[width=\hsize]{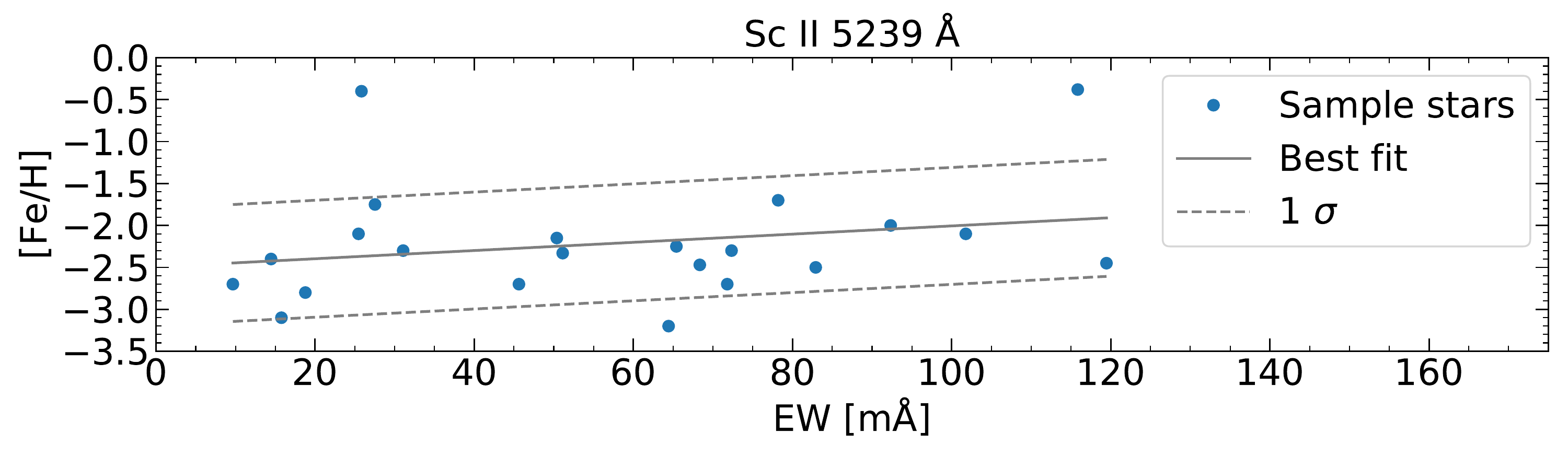}
      \caption{[Fe/H] scaled from the Sc II lines at 5239\,\AA\, detected in 22 sample stars (see Table~\ref{A-tab-ew}). We note the y-axis is larger than in  Fig. 11. A poor fit with a large scatter resulting in a residual standard deviation of 0.697\,dex is seen. }
        \label{BadSc}
   \end{figure}

 \begin{sidewaystable*}
\caption{SP\_Ace and ATHOS metallicities as well as derived metallicities with different equivalent width relations.}             
\label{tab:stellar_pars}    
\centering                     
\begin{tabular}{lrrrrrrrrr}
\hline\hline          
\noalign{\smallskip}
Object &$\mathrm{[Fe/H]}_\mathrm{ATHOS}$ & $\mathrm{[Fe/H]}_\mathrm{SP\_Ace}$ &  $\mathrm{[Cr/H]}_\mathrm{SP\_Ace}$& $\mathrm{[Ni/H]}_\mathrm{SP\_Ace}$& $\mathrm{[Fe/H]}_\mathrm{Cr,5204}$&$\mathrm{[Fe/H]}_\mathrm{Cr,5206}$&$\mathrm{[Fe/H]}_\mathrm{Cr,5208}$&$\mathrm{[Fe/H]}_\mathrm{Cr,5204+5208}$&$\mathrm{[Fe/H]}_\mathrm{Ni,5476}$ \\
\noalign{\smallskip}
\hline                  
\noalign{\smallskip}
HE0002-1037 & $-2.43$ & - & -             & -         &   -       &  -       & -       & -       & -     \\
HE0020-1741 & $-2.96$ & - & -             & -         &   -       &  -       & -       & -       & -     \\
HE0039-2635 & $-3.05$ & - & -             & -         &   -       &  -       & -       & -       & -     \\
HE0059-6540 & $-2.28$ & $-2.24$ & $-2.47$ & $-2.46$   &  $-2.57$  & $-2.56$  & $-2.47$ & $-2.60$ &$-2.45$\\
HE0221-3218  &$-0.09$ & $-0.11$  & $-0.12$& $-0.10 $  &  $-0.38$  & $-1.37$  & $0.35$  & $-0.25$ &$-0.56$\\
HE0241-3512  &$-1.39$ & $-1.51$  &  -     & $-1.66 $  &  -        & -        & -       & -       &$-0.67$\\
HE0253-6024  &$-2.03$ &  -       &  -     &  -        &  -        & -        & -       & -       & -     \\
HE0317-4705  &$-1.92$ & $-2.19$  & $-2.41$& $-2.16 $  &  $-2.43$  & $-2.50$  & $-2.37$ & $-2.49$ &$-2.15$\\
HE0400-2030  &$-2.03$ & $-2.34$  & $-2.26$& $-2.54$   &  $-2.44$  & $-2.51$  & $-2.38$ & $-2.50$ &$-2.43$\\
HE0408-1733  &$0.25 $ & $0.08 $  & $0.01 $& $0.12  $  &  $-0.30$  & $-1.33$  & $0.44$  & $-0.17$ &$-0.32$\\
HE0414-0343  &$-1.86$ & $-2.17$  & $-2.23$& $-2.19 $  &  $-2.42$  & $-2.50$  & $-2.36$ & $-2.48$ &$-2.24$\\
HE0430-4901  &$-2.60$ &  -       &  -     &  -        &  -        & -        & -       & -       & -     \\
HE0448-4806  &$-2.40$ & $-2.06$  & $-2.38$& $-2.1  $  &  $-2.88$  & $-2.70$  & $-2.76$ & $-2.87$ &$-2.71$\\
HE0516-2515  &$-0.95$ &  -       &  -     &  -        &  -        & -        & -       & -       & -     \\
HE1431-0245  &$-2.33$ & $-1.79$  &  -     & $-1.77 $  &  -        & -        & -       & -       &$-2.04$\\
HE2138-1616  &$0.20 $ & $0.21 $  & $0.19 $& $0.26  $  &  $-0.13$  & $-1.22$  & $0.72$  & $0.04$  &$-0.49$\\
HE2141-1441  &$0.33 $ & $0.14 $  & $0.07 $& $0.19  $  &  $-0.20$  & $-1.28$  & $0.58$  & $-0.06$ &$-0.32$\\
HE2144-1832  &$-0.78$ & $0.07 $  & $0.01 $& $0.16  $  &  $-0.14$  & $-1.24$  & $0.65$  & $0.01$  &$-0.27$\\
HE2153-2323  &$-1.65$ & $-1.74$  & $-2.24$&  -        &  $-2.38$  & $-2.44$  & $-2.15$ & $-2.35$ & -     \\
HE2158-5134  &$-3.14$ &  -       &  -     &  -        &  -        & -        & -       & -       & -     \\
HE2235-5058  &$-2.46$ & $-2.37$  &  -     & $-2.20 $  &  -        & -        & -       & -       &$-2.42$\\
HE2250-4229  &$-2.53$ & $-1.74$  &  -     &  -        &  -        & -        & -       & -       & -     \\
HE2258-4427  &$-2.18$ & $-2.19$  &  -     & $-2.22 $  &  -        & -        & -       & -       &$-1.98$\\
HE2310-4523  &$-2.30$ & -        &  -     &  -        &  -        & -        & -       & -       & -     \\
HE2319-5228  &$-3.05$ & -        &  -     &  -        &  -        & -        & -       & -       & -     \\
HE2339-4240  &$-2.29$ & -        &  -     &  -        &  -        & -        & -       & -       & -     \\
HE2357-2718  &$0.34 $ & $0.38 $  & $0.44 $& $0.48  $  &  $1.43$   & $-0.37$  & $2.75$  & $1.73$  &$-0.19$\\
HE2358-4640  &$-1.83$ & $-1.74$  & $-1.98$& $-1.83 $  &  $-2.46$  & $-2.51$  & $-2.39$ & $-2.51$ &$-2.12$\\

\hline                             
\end{tabular}
\end{sidewaystable*}

\end{appendix}

\end{document}